\documentclass[11pt]{article}
\pdfoutput=1
\usepackage[utf8]{inputenc}
\usepackage[english]{babel}

\usepackage{setspace}
\usepackage{simplewick}
\usepackage[dvipsnames]{xcolor}
\usepackage[normalem]{ulem}
\usepackage{cite}
\usepackage{dsfont}
\usepackage{marginnote}
\usepackage{placeins}
\usepackage{xcolor}
\usepackage{caption}
\usepackage{subcaption}

\usepackage{tikz}
\usepackage{pgfplots}
\usetikzlibrary{decorations,calligraphy}
\usetikzlibrary{decorations.pathmorphing}
\tikzset{snake it/.style={decorate, decoration=snake}}
\usetikzlibrary{calc}
\usetikzlibrary{math}
\usepackage{float}

\def\centerarc[#1](#2)(#3:#4:#5)
{ \draw[#1] ($(#2)+({#5*cos(#3)},{#5*sin(#3)})$) arc (#3:#4:#5); }

\usetikzlibrary{intersections}
\usetikzlibrary{patterns, arrows.meta, scopes, fadings, intersections, pgfplots.fillbetween}
\usepackage{blindtext}

\pgfplotsset{compat=1.18} 

\newcommand{\de}{\text{d}}

\newcommand{\xdownarrow}[1]{%
  {\left\downarrow\vbox to #1{}\right.\kern-\nulldelimiterspace}
}
\newcommand{\ve}{\varepsilon}

\usepackage{geometry}
\usepackage{verbatim}
\usepackage{prettyref}
\usepackage{amsmath}
\usepackage[normalem]{ulem}
\usepackage{amssymb}
\usepackage{graphicx}
\usepackage{setspace}
\usepackage{esint}
\usepackage{verbatim}
\usepackage{physics}
\usepackage{listings}
\usepackage{dsfont}

\usepackage{color}
\definecolor{darkgreen}{rgb}{0,0.5,0}
\definecolor{darkblue}{rgb}{0,0,0.6}
\definecolor{purple}{rgb}{0.4,.2,0.7}
\definecolor{orange}{rgb}{0.95, 0.5, 0.3}
\definecolor{gravity}{rgb}{0.88, 0.92, 1}

\usepackage[colorlinks=true,citecolor=darkgreen,linkcolor=darkblue,urlcolor=blue]{hyperref}

\numberwithin{equation}{section}
\numberwithin{table}{section}

\def\cH{{\cal H}}

\def\a{\alpha}
\def\b{\beta}

\def\cA{\mathcal A}
\def\cB{\mathcal B}
\def\cC{\mathcal C}
\def\cD{\mathcal D}

\def\cF{\mathcal F}

\def\cH{\mathcal H}

\def\cL{\mathcal L}
\def\cM{\mathcal M}
\def\cN{\mathcal N}
\def\cO{\mathcal O}

\def\cS{\mathcal S}

\def\be{\begin{equation}}
\def\ee{\end{equation}}
\def\bea{\begin{eqnarray}}
\def\eea{\end{eqnarray}}
\def\ba{\begin{align}}
\def\ea{\end{align}}

\def\cO{{\cal O}}
\def\de{\text{d}}

\def\op{\operatorname}

\definecolor{codegreen}{rgb}{0,0.6,0}
\definecolor{codegray}{rgb}{0.5,0.5,0.5}
\definecolor{codepurple}{rgb}{0.58,0,0.82}
\definecolor{backcolour}{rgb}{0.95,0.95,0.92}

\lstdefinestyle{mystyle}{
    backgroundcolor=\color{backcolour},   
    commentstyle=\color{codegreen},
    keywordstyle=\color{magenta},
    numberstyle=\tiny\color{codegray},
    stringstyle=\color{codepurple},
    basicstyle=\ttfamily\footnotesize,
    breakatwhitespace=false,         
    breaklines=true,                 
    captionpos=b,                    
    keepspaces=true,                 
    numbers=left,                    
    numbersep=5pt,                  
    showspaces=false,                
    showstringspaces=false,
    showtabs=false,                  
    tabsize=2
}

\makeatletter
\newcommand{\vast}{\bBigg@{4}}
\newcommand{\Vast}{\bBigg@{5}}
\makeatother

\begin{document}
\begin{spacing}{1.1}
  \setlength{\fboxsep}{3.5\fboxsep}

~
\vskip5mm

\begin{center}

{\Huge \textsc \textbf{The Influence Functional in open holography: \\ entanglement and Rényi entropies}}

\vskip10mm

\thispagestyle{empty}

Pietro Pelliconi \& Julian Sonner \\
\vskip1em
{\small
{\it Department of Theoretical Physics, University of Geneva \\ 24 quai Ernest-Ansermet, 1211 Genève 4, Suisse}}
\vskip5mm

\tt{pietro.pelliconi@unige.ch, julian.sonner@unige.ch}

\end{center}

\vskip10mm

\begin{abstract}

Open quantum systems are defined as ordinary unitary quantum theories coupled to a set of external degrees of freedom, which are introduced to take on the rôle of an unobserved environment. Here we study examples of open quantum field theories, with the aid of the so-called Feynman-Vernon Influence Functional (``IF"), including field theories that arise in holographic duality. We interpret the system in the presence of an IF as an open effective field theory, able to capture the effect of the unobserved environment. Our main focus is on computing Rényi and entanglement entropies in such systems, whose description from the IF, or ``open EFT", point of view we develop in this paper. The issue of computing the entanglement-Rényi entropies in open quantum systems is surprisingly rich, and we point out how different prescriptions for the IF may be appropriate depending on the application of choice. A striking application of our methods concerns the fine-grained entropy of subsystems when including gravity in the setup, for example when considering the Hawking radiation emitted by black holes. In this case we show that one prescription for the IF leads to answers consistent with unitary evolution, while the other merely reproduces standard EFT results, well known to be inconsistent with unitary global evolution. We establish these results for asymptotically AdS gravity in arbitrary dimensions, and illustrate them with explicit analytical expressions for the IF in the case of matter-coupled JT gravity in two dimensions.

\end{abstract}

\pagebreak
\setcounter{page}{1}
\pagestyle{plain}

\setcounter{tocdepth}{2}
{}
\vfill
\tableofcontents

\section{Introduction}
Holographic duality typically involves a pairing of a unitarily evolving or ``closed" quantum system, for example a quantum field theory in $d$ dimensions, with a quantum gravitational theory in $d+1$ dimensions \cite{Maldacena:1997re}. An interesting, but much less explored, scenario concerns holographic duality for open quantum systems, where we consider one or the other (or indeed both) of the members of the holographic pairing to be coupled to an external environment.  While understanding holographic open systems is by itself an interesting undertaking which we wish to advance, there are a number of concrete areas of application we wish to particularly highlight. 
 
 A first major motivation comes from considerations of black-hole physics in holographic duality, and in particular recent developments on the entropy of Hawking radiation \cite{Almheiri:2019psf,Penington:2019npb}. In these studies, AdS black holes are coupled to a bath (`environment') capable of capturing their Hawking radiation and allowing otherwise thermodynamically stable AdS black holes to evaporate. Indeed one of the main applications of the framework developed in this paper will be to determine the limitations of EFT calculations of the entropy of Hawking radiation that arise from carefully considering the dynamics of open AdS/CFT pairs.

A second motivation comes from considering recent proposals to experimentally realise holographically dual pairs in laboratory quantum systems \cite{Danshita:2016xbo,Garcia-Alvarez:2016wem,Brzezinska:2022mhj,Uhrich:2023ddx}. For example in the proposal of \cite{Uhrich:2023ddx} a high-finesse optical cavity hosts a fermionic condensate which can be engineered to implement the non-local SYK-interactions. As explained in \cite{Uhrich:2023ddx}, such a cavity-QED platform necessarily involves loss channels via the scattering of photons outside the cavity, which may be described by considering the addition of Lindblad operators to the Hamiltonian description of the system (see Appendix of \cite{Uhrich:2023ddx}). It is natural to expect that other experimental realizations must deal with their own specific loss channels and should be understood as open quantum systems in each case. This is clearly a wide and interesting field of study which we expect to receive increased attention as the required quantum technologies become available to simulate holographic systems.

The approach we adopt in our study, adapted to each application mentioned beforehand, is to employ the Feynman-Vernon influence formalism in holographic duality. In their seminal paper \cite{Feynman:1963fq}, Feynman and Vernon were the first to propose that the path integral formalism of quantum mechanics is a convenient tool in the context of open quantum systems, since the effect of the environment on the system can be accounted for by introducing a (non-local) functional which depends on the system's coordinates only. They called this functional the {\it Influence Functional}, since it encodes the influence the environment has on the system. In \cite{Feynman:1963fq}, Feynman and Vernon considered non-relativistic quantum mechanics and analyzed different examples. In particular, they were able to exactly compute the Influence Functional for an environment composed of harmonic oscillators linearly coupled to the system coordinates. On the other hand, they also showed how it is possible to perturbatively evaluate the Influence Functional for theories which are not exactly solvable. 

Subsequently, this approach proved to be very useful as perhaps first evidenced in \cite{CALDEIRA1983587} (see also \cite{CALDEIRA1983374}), nowadays referred as {\it the Caldeira-Leggett model}, where the authors gave a microscopic quantum-mechanical derivation of Brownian motion. To do so, they used Feynman's and Vernon's exact results for the Influence Functional of a system linearly coupled to a bath of harmonic oscillators. Interestingly, the authors were able to show how, choosing a specific distribution of the frequencies, the model is equivalent to an external random force acting on the system, namely a Brownian motion, which in the classical limit $\hbar \to 0$, recovers the Fokker-Planck equation. 

A main advantage of this method in the present context, is that it is clearly well adapted for relativistic Lagrangian Quantum Field Theories, which are naturally described in terms of path integrals. In Section \ref{sec:warm_up} we apply this formalism to simple Quantum Field Theories, to show how it is possible to systematically obtain the Influence Functional. This is instrumental for introducing the fundamental concepts that will be used for holographic theories, which also lend themselves well to a Lagrangian-type analysis via the path integral.

\begin{figure}[t]
\begin{center}
\begin{tikzpicture}

\draw[very thick] plot[domain=0:1, smooth, samples=20, variable=\t] ({\t-1}, {0.05*sin(1000*\t)});

\fill[Cerulean, opacity=0.2] (-2.25,0) circle (1.25);
\draw[Cerulean, very thick] (-2.25,0) circle (1.25);

\fill[Goldenrod, opacity=0.2] (1.25,0) circle (1.25);
\draw[Goldenrod!50!BurntOrange, very thick] (1.25,0) circle (1.25);

\draw[Goldenrod!50!BurntOrange] (1.25, 0.25) node{Environment};
\draw[Goldenrod!50!BurntOrange] (1.25, -0.25) node{$\chi$};

\draw[Cerulean] (-2.25, 0.25) node{System};
\draw[Cerulean] (-2.25, -0.25) node{$\phi$};

\draw[] (3.5, 0) node{\LARGE $=$};

\begin{scope}[shift = {(8, 0)}]

\fill[Goldenrod, opacity=0.2] (-2.15,0) circle (1.25);
\draw[Goldenrod!50!BurntOrange, very thick] (-2.15,0) circle (1.25);

\fill[Cerulean, opacity=0.2] (-2.35,0) circle (1.25);
\draw[Cerulean, very thick] (-2.35,0) circle (1.25);

\draw[ForestGreen] (-2.25, 0.30) node{Open EFT};
\draw[ForestGreen] (-2.25, -0.30) node{$\phi \, $,  $\, \cF[\phi]$};
\end{scope}

\end{tikzpicture}
\end{center}
\caption{A pictorial description of a generic open quantum system, where the system under consideration is coupled to an environment. In the Lagrangian path integral formalism, the effect the environment has on the system can be taken into account introducing a functional, the Influence Functional, to the system path integral. Remarkably, this functional depends (non-locally) only on the system degrees of freedom. We refer to the resulting effective theory as Open EFT.}
\label{it}
\end{figure}
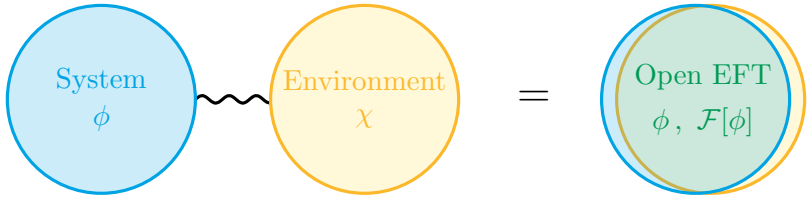

Our main goal is thus to develop a framework for open AdS/CFT in general terms, that is a theory of system-environment interaction valid for arbitrary holographic pairs. We use the Feynman-Vernon Influence Functional as the starting point for this development, since it appears particularly well suited to the holographic dictionary as it formulates the dynamics of open systems in a Lagrangian, rather than Hamiltonian, framework. The basic idea \cite{Feynman:1963fq} is to formulate the system-environment interplay in terms of the total path integral
\begin{equation}\label{eq.InfluenceFunctionalIntro}
\int [{\cal D}\phi ]  [{\cal D}\chi ] \,  e^{-I_{\rm sys}[\phi] - I_{\rm env}[\chi] - I_{\rm c} [\phi, \chi]} = \int [{\cal D}\phi ] \, {\cal F}[\phi] \, e^{-I_{\rm sys}[\phi]}\,,
\end{equation}
where ${\cal F}[\phi]$ is the -- typically non-local -- Influence Functional of Feynman and Vernon\footnote{Strictly speaking, Feynman and Vernon defined their Influence Functional with two insertions of the state of the system ${\cal F}^{\rm FV}[\phi,\phi']$. This is instrumental if one wants to consider out-of-equilibrium dynamics, because it generates the Schwinger-Keldysh contour. In this work, we will consider equilibrium states which also have a well defined Euclidean description. Therefore, we will content ourselves with computing the Influence Functional of \eqref{eq.InfluenceFunctionalIntro}.}. The system and environment setup envisaged here is illustrated schematically in Figure \ref{it}, and we generically denote system fields as $\phi$ and environment fields as $\chi$. Of course at this formal level the procedure amounts to little more than path-integrating out the environment, but the practical advantages become apparent when one considers realistic situations where one only has access to the system degrees of freedom, which are however coupled to a potentially unknown environment. In such cases, the Influence Functional is simply interpreted as a property of the system itself and the resulting description acquires the status of an Effective Field Theory description. As we will show in this work, thinking in this way gives new insight into computations of entanglement entropy of open systems, and is particulary striking in the gravitational context. 

The Feynman-Vernon Influence Functional approach has been developed holographically in previous works. For example \cite{Jana:2020vyx,Loganayagam:2022zmq} studied the problem of open-system dynamics with respect to a strongly coupled bath. In these works an Open EFT is derived on the Schwinger-Keldysh contour, relevant for non-equilibrium (e.g. OTO) correlation functions, without including non-trivial `replica-saddles'. In this work we instead focus on entanglement and Rényi entropies, which means that we study the Euclidean theory, but we do include the contribution of non-trivial `replica-saddles'. Extending the analysis of this paper to Schwinger-Keldysh type contours, as was done in \cite{Jana:2020vyx,Loganayagam:2022zmq} for local correlation functions, would be of essential interest in addressing the issue of entanglement and related quantities in non-equilibrium situations. This constitutes an interesting problem to be addressed in the future.

The language of the Influence Functional gives a new perspective on the system under consideration. While correlation functions of operators in the system give identical results both in the original theory and using the Influence Functional, this is not necessarily true for more fine grained observables like the entanglement entropy, as it is famously known for Quantum Mechanical theories. In particular, in Section \ref{sec:warm_up} below we show how it is possible to compute the entanglement entropy of the reduced system density matrix using the Influence Functional and the Open EFT. The same machinery can also be applied to the EFT of gravity coupled to matter, to compute the matter entanglement entropy. In doing so we encounter one of the various incarnations of the information paradox. Thus, as an example of perhaps surprisingly diverse applications of the topic of open quantum system, this gives us a different perspective on the applicability of semiclassical EFT of gravity coupled to matter to compute the entanglement entropy of Hawking's radiation, and in general of gravitational theories. Our proposed solution, namely a prescription to compute the gravitational Influence Functional for the replicated theory, from the point of view of open quantum systems raises even more questions on the interplay between quantum matter and holographic systems. We summarize possible interpretations and future intriguing directions in the discussion Section \ref{sec:Discussion}.

The plan of the paper is as follows. In Section \ref{sec:warm_up} we introduce the Influence Functional in more detail, exhibiting some generic properties, and computing it for simple example theories where we show how to systematically compute it perturbatively. We then compute entanglement entropies for simple theories, namely free massive scalars in two dimensions. In particular, we show how to compute the reduced entanglement entropy of a system linearly coupled to an environment composed of a free massive scalar. In Section \ref{sec:Hawking_rad_EE}, we use the same machinery to compute the Influence Functional of a scalar field minimally coupled to gravity in AdS$_d$, that is we move into the arena of holography. Treating gravity as an EFT, we use this Influence Functional to show that computing the reduced matter entanglement entropy results, at leading order, to the entanglement entropy of matter field on the fixed gravitational background. 
We then propose a different prescription than the `naive' EFT approach, when using the replica trick, which is able to reproduce a matter entanglement entropy consistent with global unitarity. We apply this machinery in Section \ref{JT_gravity_and_isl_model} to JT-gravity coupled to a CFT$_2$, in the so-called {\it island model}. Last, in Section \ref{sec:Discussion} we comment on the results found, arguing why the prescription proposed to compute the Influence Functional, even though it has a semi-classical flavor, is rather unnatural from an EFT perspective. We then suggest how these results can potentially reveal interesting microscopic properties of holographic theories. In particular, our findings fit well with the idea that an underlying coarse graining is needed to obtain a geometric theory of gravity. There is mounting of evidence for this being true in some low-dimensional examples, like SYK-JT gravity \cite{Saad:2019lba,Altland:2022xqx}, or more recently AdS$_3$/CFT$_2$ \cite{Belin:2020hea,Chandra:2022bqq,Belin:2023efa}, but unknown for higher dimension. In Appendix \ref{app:Coarse_grain_EE} we show the connection between our findings and a coarse graining in entanglement entropies.

\section{Influence Functionals and entanglement entropies} \label{sec:warm_up}

The basic idea of analysing an open quantum system with the Influence Functional approach is quite simple. As was briefly mentioned in the introduction, we start out by considering a system that is coupled to an environment, but such that the microscopic theories describing the two are assumed to be known, as well as the coupling between them, calling the system degrees of freedom $\phi$ and the environment degrees of freedom $\chi$, as in \eqref{eq.InfluenceFunctionalIntro}. Because of the coupling, the dynamics of the system is modified by the environment, in such a way that in the system's path integral there appears a functional, ${\cal F}$, the so-called Influence Functional. Let us now show this more explicitly. The global system has action
\begin{equation}
	I[\phi, \chi] = I_{\rm sys} (\phi)  +  I_{\rm env} (\chi)  + I_{\rm c} (\phi, \chi)  \ ,
	\label{action_generic}
\end{equation}
where $I_{\rm c}(\phi, \chi)$ is the term in the action that contains all interactions between system and environment. The partition function of the global system is thus
\begin{equation}
	Z = \int [\cD \phi][\cD \chi] \, e^{- I_{\rm sys} (\phi) - I_{\rm env} (\chi) - I_{\rm c} (\phi, \chi)} \ .
	\label{original_theory}
\end{equation}
If we want to find the effective theory for the system we need to isolate the environment degrees of freedom
\begin{equation}
	Z = \int [\cD \phi] \left( \int [\cD \chi] \, e^{ - I_{\rm env} (\chi) - I_{\rm c} (\phi, \chi)} \right) \, e^{- I_{\rm sys} (\phi)} =  \int [\cD \phi] \, \cF[\phi] \, e^{- I_{\rm sys} (\phi)} \ .
	\label{Reduced_Theory_with_IF}
\end{equation}
The functional $\cF[\phi]$ is the Influence Functional, and it is defined as 
\begin{equation}
	 \cF[\phi] = \int  [\cD \chi] \, e^{ - I_{\rm env} (\chi) - I_{\rm c} (\phi, \chi)} \ .
	 \label{IF_Generic_Expression}
\end{equation}
In general it is a functional in the system's degrees of freedom $\phi$, which encodes all the information on the environment and how it modifies the dynamics of the system. In some special cases it can be computed exactly, while generally one can only compute a perturbative approximation, which can be obtained systematically using standard Feynman diagram techniques\footnote{In many applications, moreover, such a perturbative evaluation is done on a Schwinger-Keldysh contour, allowing one to address non-equilibrium problems.}. 

Notice also that in general the Influence Functional \eqref{IF_Generic_Expression} is non-local in time, which induces a non-Markovian evolution for the system's degrees of freedom.  In order to gain some intuition, in the next Section we will compute the Influence Functional in some very simple examples,  such as the case of a linear coupling between system and environment.

\subsection{A simple example}

Consider the action
\begin{equation}
	I[\phi, \chi] = I_{\rm sys} [\phi] + I_{\rm env} [\chi] + \eta \, \int \de^d x \, \phi(x) \, \chi(x)   \ ,
	\label{Generic_action_interaction}
\end{equation}
where the environment is possibly not a free theory. We compute the Influence Functional with \eqref{IF_Generic_Expression}. Expanding perturbatively in $\eta$ we get
\begin{equation}
	\cF = \sum_{l = 0}^{\infty} \frac{(-1)^l \, \eta^l }{l!} \int \de x_1 \dots \de x_l \, \phi(x_1) \dots \phi(x_l) \, \int [\cD \chi] \, e^{- I_{\rm env}[\chi] } \,  \chi(x_1) \dots  \chi(x_l)  \ .
\end{equation}
This can of course simply be rewritten as 
\begin{equation}
	\cF = Z_{\chi} \sum_{l = 0}^{\infty} \frac{(-1)^l \, \eta^l }{l!} \int \de x_1 \dots \de x_l \, \phi(x_1) \dots \phi(x_l)  \,  G_{\chi}(x_1, \dots  , x_l)  \ ,
	\label{IF_interacting_before_resum}
\end{equation}
where
\begin{equation}
	Z_{\chi} =  \int [\cD \chi] \, e^{- I_{\rm env}[\chi] } 
\end{equation}
is the partition function of the environment, while $G_{\chi}(x_1, \dots  , x_n)$ are its $n$-point correlation functions, namely 
\begin{equation}
	G_{\chi} (x_1, \dots , x_n) = \frac{1}{Z_{\chi}} \int [\cD \chi] \, \chi(x_1) \dots \chi(x_n) \, e^{- I_{\rm env}[\chi] } \ ,
\end{equation}
containing both connected and non-connected diagrams. The series \eqref{IF_interacting_before_resum} can be re-summed \cite{Zinn-Justin:572813} into 
\begin{equation}
	\cF = Z_{\chi} \exp \left( \sum_{l = 1}^{\infty} \frac{(-1)^l \, \eta^l }{l!} \int \de x_1 \dots \de x_l \, \phi(x_1) \dots \phi(x_l)  \,  G_{\chi}^{c}(x_1, \dots  , x_l) \right) \ ,
	\label{IF_with_Interactions}
\end{equation}
where $G_{\chi}^{c}(x_1, \dots  , x_n)$ comprises only connected diagrams, hence the superscript. All in all, the effective theory for the system degrees of freedom of \eqref{Generic_action_interaction} becomes 
\begin{equation}
	Z = Z_{\chi} \int [\cD \phi] \, \exp\left(- I_{\rm sys} (\phi) + \sum_{l = 1}^{\infty} \frac{(-1)^l \, \eta^l }{l!} \int \de x_1 \dots \de x_l \, \phi(x_1) \dots \phi(x_l)  \,  G_{\chi}^{c}(x_1, \dots  , x_l) \right) \ ,
	\label{Reduced_Theory_Linear_Coupling_with_interactions}
\end{equation}
which, as expected, is non-local. We will usually call the partition function written in this fashion the {\it open effective field theory\footnote{In analogy with the term `open effective action' used previously in \cite{Jana:2020vyx}. }}, since it is a description purely in term of the system's degrees of freedom. Note, however, that EFTs typically involve a mass scale that suppresses non-local interactions and higher-dimension operators. In our Open EFT we do not necessarily have such a mass scale, for example in the interesting case of gravity which we treat below. In the simpler examples of massive scalar theories of Section \ref{sec:QFT_EE}, the mass of the scalar that is being integrated out could be taken to furnish such a scale, for example by organising \eqref{Reduced_Theory_Linear_Coupling_with_interactions} in a derivative expansion with higher-order suppressed by inverse powers of the mass. Note, however, that we do not assume any parametric separation of the mass of the system and that of the environment (in fact we take the two to be the same), so that strictly speaking even in this example the Open EFT is never fully local, as remarked before.

This effective theory can be used to study all observables of the system. For example, we can use it to compute correlation functions, and the result using the original theory exactly matches the one using the Influence Functional. This is trivial to show,
\begin{align}
	\langle \phi(x_1) \dots \phi(x_n) \rangle \, = \, & \frac{1}{Z} \int [\cD \phi][\cD \chi] \,  \phi(x_1) \dots \phi(x_n) \, e^{- I_{\rm sys} [\phi] - I_{\rm env} [\chi] - I_{\rm c} [\phi, \chi] } \nonumber \\
	= \, & \frac{1}{Z} \int [\cD \phi] \,  \phi(x_1) \dots \phi(x_n) \, \cF[\phi] \,  e^{- I_{\rm sys} [\phi]} \ .
\end{align}
The Influence Functional is, of course, nevertheless a highly useful device as it allows to abstractly describe new features that appear in the dynamics of quantum systems coupled to environments.
In the following we will be mostly interested in a different observable, namely the entanglement entropy. In particular, computing it from effective theories of the form \eqref{Reduced_Theory_Linear_Coupling_with_interactions}, one expects as a result to obtain the entanglement entropy of the system's degrees of freedom after tracing out the environment. However, unlike correlation functions, entanglement (and R\`enyi) entropies require more care, and different answers can result from different prescriptions of integrating out the environment, as we will show in this paper. Naturally, we do not claim that the results obtained from the Influence Functional are ambiguous, merely that we have a number of choices that can be made, according to the situation under consideration, that affect the precise form of the Influence Functional that appears when computing entanglement and R\`enyi entropies.

\subsection{Entanglement entropy for EFTs and the replica trick} \label{sec:Enta_Entr}

We will be interested in computing entanglement entropies for open systems, both in the context of Open EFT and in gravitational theories. To this end, we will review some basic constructions in this subsection before moving on to implement these in the Influence Functional approach.

Given a density matrix $\rho$ describing a quantum system, the entanglement entropy is defined as
\begin{equation}
	S(\rho) = - \op{Tr} [\rho \log(\rho)] \ .
    \label{EE_definition}
\end{equation}
Our focus will be to study the entanglement entropies both in open non-gravitational quantum field theories as well as in gravitational examples. We study the former in this section and the latter in the next one. In both cases, we are interested in the entanglement entropy of the system's degrees of freedom after tracing out the environment. This can be naturally described using the Influence Functional. As a technical remark, we always work in Euclidean signature.

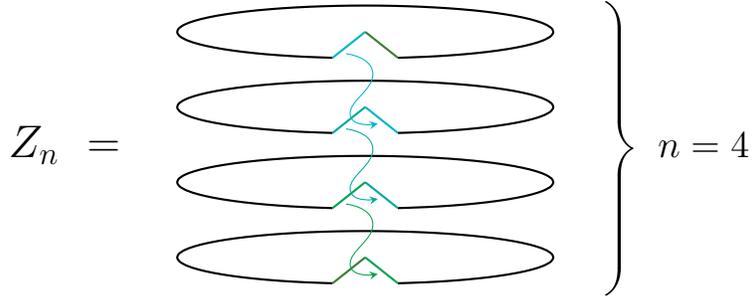
\begin{figure}[t]
\centering
\begin{tikzpicture}

\tikzset{
    partial ellipse/.style args={#1:#2:#3}{
        insert path={+ (#1:#3) arc (#1:#2:#3)}
    }
}

\draw[thick] plot[domain=-80:260, smooth, samples=40, variable=\x] ({2.5*cos(\x)}, {0.35*sin(\x)});
\draw[thick, OliveGreen] (0, 0) -- ({2.5*cos(-80)}, {0.35*sin(-80)});
\draw[thick, Turquoise] ({2.5*cos(260)}, {0.35*sin(260)}) -- (0, 0);

\draw[thick] plot[domain=-80:260, smooth, samples=40, variable=\x] ({2.5*cos(\x)}, {-1 + 0.35*sin(\x)});
\draw[thick, Turquoise] (0, -1) -- ({2.5*cos(-80)}, {-1 + 0.35*sin(-80)});
\draw[thick, Emerald] ({2.5*cos(260)}, {-1 + 0.35*sin(260)}) -- (0, -1);

\draw[thick] plot[domain=-80:260, smooth, samples=40, variable=\x] ({2.5*cos(\x)}, {-2 + 0.35*sin(\x)});
\draw[thick, Emerald] (0, -2) -- ({2.5*cos(-80)}, {-2 + 0.35*sin(-80)});
\draw[thick, Green] ({2.5*cos(260)}, {-2 + 0.35*sin(260)}) -- (0, -2);

\draw[thick] plot[domain=-80:260, smooth, samples=40, variable=\x] ({2.5*cos(\x)}, {-3 + 0.35*sin(\x)});
\draw[thick, Green] (0, -3) -- ({2.5*cos(-80)}, {-3 + 0.35*sin(-80)});
\draw[thick, OliveGreen] ({2.5*cos(260)}, {-3 + 0.35*sin(260)}) -- (0, -3);

\draw[Turquoise, -stealth] plot [smooth, tension=1, ] coordinates { ({1.75*cos(260) + 0.05}, {0.245*sin(260) - 0.05}) (0.1, -0.55)  (-0.2, -1.10) ({1.25*cos(-80) - 0.05}, {-1 + 0.175*sin(-80) - 0.05}) };

\draw[Emerald, -stealth] plot [smooth, tension=1, ] coordinates { ({1.75*cos(260) + 0.05}, {0.245*sin(260) - 0.05 - 1}) (0.1, -1.55)  (-0.2, -2.10) ({1.25*cos(-80) - 0.05}, {0.175*sin(-80) - 0.05 - 2}) };

\draw[Green, -stealth] plot [smooth, tension=1, ] coordinates { ({1.75*cos(260) + 0.05}, {0.245*sin(260) - 0.05 - 2}) (0.1, -2.55)  (-0.2, -3.10) ({1.25*cos(-80) - 0.05}, {0.175*sin(-80) - 0.05 - 3}) };

\draw[black] (4.5, -1.5) node{\Large $n=4$};

\draw [very thick,decorate,decoration={calligraphic brace,amplitude=10pt}] (3.2, 0.4) -- (3.2,-3.5);

\draw[black] (-4, -1.5) node{\LARGE $Z_n \, \, =$};
\end{tikzpicture}
\caption{The $n$-branched replica manifold used to compute the entanglement entropy of a (semi-infinite) interval via the replica thick. We show $n = 4$ as an example. Every level is connected to the previous and the next one through the branch cuts, as drawn. The arrow connecting the two dark green lines is not shown for illustration purposes. }
\label{fig:Replica_trick}
\end{figure}

Entanglement entropies in Effective Field Theories are traditionally computed by employing the {\it replica trick}. This procedure entails computing first the Rényi entropies
\begin{equation}
	S_n(\rho) = \frac{1}{1-n} \log \Big( \op{Tr} [\rho^n] \Big)  \ ,
	\label{EE_replica_density_matrix}
\end{equation}
and then relating them to the entanglement entropy via the limit
\begin{equation}
	S(\rho) = \lim_{n \to 1} S_n(\rho) \ .
\end{equation}
This procedure works as long as the replica symmetry $\mathbb Z_n$ is not broken, and it will always be the case in the following. It turns out that for Quantum Field Theories employing the replica trick is simpler than diagonalizing the (reduced) density matrix of the state considered. The reason is that it is possible to write a density matrix in terms of the partition function of the system on a background that prepares the state. More precisely, if we have a manifold containing boundaries which are complete Cauchy slices, we can specify states over those boundaries. Then, the Euclidean evolution over such manifold can be seen as a (un-normalized) state preparation, 
\begin{equation}
	\rho = e^{- (\tau_2 - \tau_1)H}/Z \ ,
\end{equation}
and thus by definition
\begin{equation}
	Z\big[ \phi(\tau_1) \to \phi(\tau_2) \big] = \bra{\phi_2} e^{- (\tau_2 - \tau_1)H} \ket{\phi_1}  \ .
	\label{part_funct_and_dens_matrix}
\end{equation}
A well known example is flat Euclidean space, which prepares the vacuum, since the evolution operator becomes a projector to that state. However, in general any manifold through a Euclidean evolution prepares a (unnormalized) density matrix. This becomes very handy when using the replica trick, since we can connect computing moments of $\rho$ to computing partitions function over the state-preparation manifold. For instance, a state $\rho$ is specified with a cut in the manifold, so that the moments $\op{Tr} [\rho^n]$ are computed through the partition $Z_n$ function on a geometry that links all the replicas connecting one side of the cut to the one of the cuts of the previous or the next replica sheet. As an example, Figure \ref{fig:Replica_trick} shows the replicated manifold when computing the entanglement entropy of a (semi-infinite) interval in a two-dimensional EFT in the vacuum. The upshot is that the entanglement entropy can be obtained by
\begin{equation}
	S = \lim_{n \to 1} \frac{1}{1-n} \log \left( \frac{Z_n}{Z^n} \right) \ .
	\label{EE_replica_trick}
\end{equation}
The denominator $Z^n$, which is the $n$-th power of the partition function of the non-replicated geometry, takes care of the normalisation in the state preparation.

In the next section we will explain how the Influence Functional fits into this picture.

\subsection{Influence Functional and the replicated geometry}

We now want to apply this technology to open quantum systems, described in the language of Influence Functionals. In particular, we have seen at the beginning of Section \ref{sec:warm_up} how we can take into account the effect of the environment on the system with a (generically non-local) functional in the system's path integral. Let us repeat it here for convenience. The system's effective dynamics can be described by
\begin{equation}
    Z = \int [\cD \phi] \, e^{- I_{\rm sys} (\phi)} \, \cF[\phi]\ ,
    \tag{\ref{Reduced_Theory_with_IF}}
\end{equation}
where 
\begin{equation}
    \cF[\phi] = \int  [\cD \chi] \, e^{ - I_{\rm env} (\chi) - I_{\rm c} (\phi, \chi)} \ .
    \tag{\ref{IF_Generic_Expression}}
\end{equation}
We wish to compute the entanglement entropy of the system's degrees of freedom using the system's effective theory \eqref{Reduced_Theory_with_IF}. To do this, we have to compute $Z_n$ in the replicated manifold, and clearly we have to extend the Influence Functional to this geometry. We propose two ways in which this extension can be done, which lead to quantitatively different results. 

The key idea is that the replicated manifold is determined by the boundary conditions we impose on the fields at the cuts. Let us call $\cM^n$ the replicated manifold, and $\cM_i$ the various sheets composing it, which all have the same topology as the original $\cM$ and
\begin{equation}
    \cM^n = \bigcup_{i = 1}^n \cM_i \ .
    \label{Union_of_sheets}
\end{equation}
The various $\cM_i$ are connected though branch cuts $\cB_i$ that extend between codimension-2 surfaces. Performing a full clockwise revolution on the two-dimensional sub-manifold orthogonal to these surfaces, we pass from the sheet $\cM_i$ to the one $\cM_{i+1}$. By construction, at the position of this surfaces the background metric of $\cM^n$, which we can call $\hat g^n_{\mu \nu}$, presents conical defects of opening angle $2 \pi (n - 1)$. One example of such a manifold is the one of Figure \ref{fig:Replica_trick}, which is used to compute the entanglement entropy of an interval in the vacuum.

In order to extend the Influence Functional to this replicated geometry we need to specify the boundary conditions at the various branch cuts of the different sheets. We resolve the branch cuts considering $\cB_i^{\pm}$, where the subscript indicates adding (or subtracting) a small positive quantity $\ve$ orthogonal to the branch cut. For the system degrees of freedom, the only natural boundary condition is
\begin{equation}
    \phi_i(\cB_i^-) = \phi_{i+1}(\cB_{i+1}^+) \ .
    \label{EFT_matter_BC}
\end{equation}
The indices $i$ indicate the fields in the various sheets. If the replica symmetry is not broken, the boundary condition \eqref{EFT_matter_BC} can be used to extend the $n$ system fields $\phi_i$ to a unique field $\phi$ living on the whole spacetime $\cM^n$. In the rest of the work we will always assume this to be the case, and thus we will leave implicit the indices $i$ for the system fields. On the other hand, for the environment degrees of freedom there are two distinct possibilities.
\begin{enumerate}
    \item[A.] The environment fields on different sheets do not communicate. Each sheet has its own environment field $\chi_i$, and the boundary condition can be expressed as
    \begin{equation}
        \chi_i(\cB_i^-) =  \chi_i(\cB_{i}^+) \ . 
        \label{A_environment_BC}
    \end{equation}
    \item[B.] The environment fields of different sheets do communicate. In this case we have effectively only one environment spanning the whole $\cM^n$, and such that
    \begin{equation}
        \chi_i(\cB_i^-) =  \chi_{i+1}(\cB_{i+1}^+) \ .
        \label{B_environment_BC}
    \end{equation}
\end{enumerate}

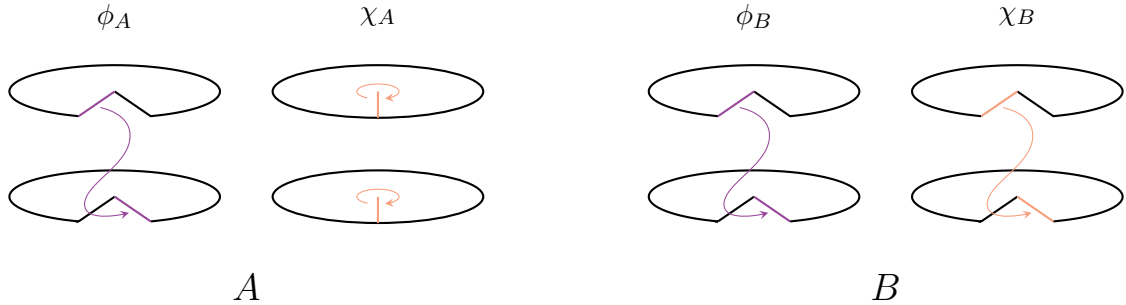
\begin{figure}[t]
\centering
\begin{tikzpicture}

\tikzset{
    partial ellipse/.style args={#1:#2:#3}{
        insert path={+ (#1:#3) arc (#1:#2:#3)}
    }
}

\draw[thick] plot[domain=-70:250, smooth, samples=40, variable=\x] ({1.4*cos(\x)}, {-1.4 + 0.35*sin(\x)}) -- ({1.4*cos(250)}, {-1.4 + 0.35*sin(250)}) -- (0, -1.4);

\draw[thick, Purple] (0, -1.4) -- ({1.4*cos(-70)}, {-1.4 + 0.35*sin(-70)});

\draw[Purple, -stealth] plot [smooth, tension=1, opacity = 0.3] coordinates { ({0.7*cos(250) + 0.05}, {0.175*sin(250) - 0.05}) (0.2, -0.65)  (-0.4, -1.50) ({0.7*cos(-70) - 0.05}, {-1.4 + 0.175*sin(-70) - 0.05}) };

\draw[thick] (0, 0) -- ({1.4*cos(-70)}, {0.35*sin(-70)}) -- plot[domain=-70:250, smooth, samples=40, variable=\x] ({1.4*cos(\x)}, {0.35*sin(\x)});

\draw[thick, Purple] ({1.4*cos(250)}, {0.35*sin(250)}) -- (0, 0);
\draw[thick] (3.5, 0) ellipse (1.4 and 0.35);
\draw[thick] (3.5, -1.4) ellipse (1.4 and 0.35);
\draw[thick, Melon] (3.5, 0) -- (3.5, -0.35);
\draw[thick, Melon] (3.5, -1.4) -- (3.5, -1.75);
\draw[Melon, -stealth] (3.5,0) [partial ellipse=240:-70: 0.28 and 0.1];
\draw[Melon, -stealth] (3.5,-1.4) [partial ellipse=240:-70: 0.28 and 0.1];

\draw[thick] plot[domain=-70:250, smooth, samples=40, variable=\x] ({1.4*cos(\x) + 8.5}, {-1.4 + 0.35*sin(\x)}) -- ({1.4*cos(250) + 8.5}, {-1.4 + 0.35*sin(250)}) -- (8.5, -1.4);
\draw[thick, Purple] (8.5, -1.4) -- ({1.4*cos(-70)+ 8.5}, {-1.4 + 0.35*sin(-70)});
\draw[Purple, -stealth] plot [smooth, tension=1, opacity = 0.3] coordinates { ({0.7*cos(250) + 0.05 + 8.5}, {0.175*sin(250) - 0.05}) (8.7, -0.65)  (8.1, -1.50) ({0.7*cos(-70) - 0.05 + 8.5}, {-1.4 + 0.175*sin(-70) - 0.05}) };
\draw[thick] (8.5, 0) -- ({1.4*cos(-70) + 8.5}, {0.35*sin(-70)}) -- plot[domain=-70:250, smooth, samples=40, variable=\x] ({1.4*cos(\x) + 8.5}, {0.35*sin(\x)});
\draw[thick, Purple] ({1.4*cos(250) + 8.5}, {0.35*sin(250)}) -- (8.5, 0);

\draw[thick] plot[domain=-70:250, smooth, samples=40, variable=\x] ({1.4*cos(\x) + 12}, {-1.4 + 0.35*sin(\x)}) -- ({1.4*cos(250) + 12}, {-1.4 + 0.35*sin(250)}) -- (12, -1.4);
\draw[thick, Melon] (12, -1.4) -- ({1.4*cos(-70)+ 12}, {-1.4 + 0.35*sin(-70)});
\draw[Melon, -stealth] plot [smooth, tension=1, opacity = 0.3] coordinates { ({0.7*cos(250) + 0.05 + 12}, {0.175*sin(250) - 0.05}) (12.2, -0.65)  (11.6, -1.50) ({0.7*cos(-70) - 0.05 + 12}, {-1.4 + 0.175*sin(-70) - 0.05}) };
\draw[thick] (12, 0) -- ({1.4*cos(-70) + 12}, {0.35*sin(-70)}) -- plot[domain=-70:250, smooth, samples=40, variable=\x] ({1.4*cos(\x) + 12}, {0.35*sin(\x)});
\draw[thick, Melon] ({1.4*cos(250) + 12}, {0.35*sin(250)}) -- (12, 0);

\draw (0, 1) node{$\phi_A$};
\draw (3.5, 1) node{$\chi_A$};
\draw (8.5, 1) node{$\phi_B$};
\draw (12, 1) node{$\chi_B$};

\draw (1.75, -2.6) node{\Large $A$};
\draw (10.25, -2.6) node{\Large $B$};
\end{tikzpicture}
\caption{Different boundary condition we can assign to the fields in order to extend the Influence Functional to the replicated geometry. In both cases the system fields living in different replicas are coupled. On the other hand, in scenario $A$ we don't couple the environment's replicas, while we do in scenario $B$. This gives different Influence Functionals.} 
\label{fig:BC_Influence_Funcional}
\end{figure}

In both cases, it is possible that new saddles appear in the replicated geometry. To find $Z_n$ we then sum over all possible saddle-point contributions.

These two boundary conditions are pictorially represented in Figure \ref{fig:BC_Influence_Funcional}. From an open system perspective, the boundary condition \eqref{A_environment_BC} may seem, at first sight, more natural than \eqref{B_environment_BC}, since one usually assumes not to have any control over degrees of freedom which have been integrated out. We will show however that this is not always the case, and there are situations in which Scenario $B$ is better suited.
In this paper we therefore consider both kinds of boundary conditions. It may already be evident to the Reader that one of the conclusions we arrive at is that the procedures of integrating out into the Influence Functional and replicating the theory do not commute, leading to interesting physical effects for the behavior of Open EFTs. We will return to these more conceptual points later on. In the next Section we will analyze a simple example to show how these two prescription quantitatively differ. This should be seen as a warm-up exercise to proceed to more non-trivial examples, including holographic gravity.

\subsection{Entanglement entropy from the Influence Functional} \label{sec:QFT_EE}

As a first example, we apply the machinery proposed above to the case of free gapped theories in two dimensions. At first, let us consider the action of a free real massive scalar of mass $m$,
\begin{equation}
	I[\phi] = \int \de^2 x \left( \frac{1}{2} \, \partial_{\mu} \phi \, \partial^{\mu} \phi + \frac{1}{2} \, m^2 \,  \phi^2 \right) \ ,
	\label{2D_massive_scalar}
\end{equation}
The simplest entanglement entropy we can compute for this theory is the entanglement entropy of a semi-infinite interval starting from the origin and going to infinity in the vacuum. The replicated geometry is then similar to the ones presented in Figure \ref{fig:Replica_trick}, and the results is known \cite{Calabrese:2004eu} to be 
\begin{equation}
	S = \frac{1}{6} \log \left( \frac{1}{m \, \ve}\right) \ .
	\label{Entropy_for_massive_scalar_theory}
\end{equation}
where $\ve$ is a UV regulator needed in order to obtain a finite answer. One way to interpret the result \eqref{Entropy_for_massive_scalar_theory} is the following. In a gapped theory we have a correlation length of the order $\sim 1/m$, and we expect the entanglement to ``spread'' only up to scales of the order of the correlation length. For this reason, if the interval is much bigger than the correlation length we expect a result similar to \eqref{Entropy_for_massive_scalar_theory}, up to corrections due to the finite size of the interval. Likewise, the dominant contribution for the multi-interval case can be approximated as
\begin{equation}
	S \approx  \frac{\cA}{6} \, \log \left( \frac{1}{m \, \ve}\right) \ ,
\end{equation}
where $\cA$ is the number of boundary points of the union of the intervals. This is a realization of the area scaling of entanglement entropies in QFTs, and \eqref{Entropy_for_massive_scalar_theory} fits into this picture, since it has only one boundary point.

This result can be used to extract interesting conclusions about entanglement entropies for interacting theories. As a simple example, we can consider a pair of massive scalars that are linearly coupled, with action
\begin{equation}
	I[\phi, \chi] = \int \de^2 x \left( \frac{1}{2} \, \partial_{\mu} \phi \, \partial^{\mu} \phi + \frac{1}{2} \, m^2 \,  \phi^2 + \frac{1}{2} \, \partial_{\mu} \chi \, \partial^{\mu} \chi + \frac{1}{2} \, m^2 \,  \chi^2 + \eta \, \phi \, \chi \right) \ .
	\label{2D_massive_scalars_interacting}
\end{equation}
This theory is quite interesting for us because it is solvable and because the Influence Functional can be computed exactly. Indeed, this theory can be diagonalized into a pair of free decoupled scalars, with masses
\begin{equation}
    m^2_{\pm} = m^2 \pm \eta \ .
\end{equation}
Then, using the result \eqref{Entropy_for_massive_scalar_theory}, it's simple to show that the entanglement entropy for this theory is
\begin{equation}
	S_{\phi, \chi}  = \frac{1}{6} \log ( \frac{1}{m_+ \,  \ve} ) + \frac{1}{6} \log ( \frac{1}{m_- \, \ve} ) \ ,
	\label{Entropy_for_interacting_massive_scalar_theory}
\end{equation}
which is the exact result for the action \eqref{2D_massive_scalars_interacting} of the global system. We can also expand it perturbatively in $\eta/m^2$ obtaining
\begin{equation}
	S_{\phi, \chi} = \frac{1}{3} \log \left(  \frac{1}{m \,  \ve} \right)+ \frac{\eta^2}{12 m^4} + \cO (\eta^4) \ .
    \label{Entropy_for_interacting_massive_scalar_theory_perturbatively}
\end{equation}
An interesting way to write this result is 
\begin{align}
	S_{\phi, \chi} =  S_{\phi} + S_{\chi} + \frac{\eta^2}{12 m^4} + \cO (\eta^4) \ ,
	\label{Entropy_interpretation}
\end{align}
which shows perturbatively that the role of the interactions is to shift the entanglement entropy for the decoupled theory, which is the sum of the two contributions coming from the fields $\phi$ and $\chi$, both of the form \eqref{Entropy_for_massive_scalar_theory}.

We now want to compare this result with the ones obtained using the Influence Functional. In the case of the action \eqref{2D_massive_scalars_interacting}, integrating out the environment is simple. One can either perform the Gaussian path integral, or use \eqref{IF_interacting_before_resum}. Both procedures of course produce the same result, which is
\begin{equation}
	\cF = Z_{\chi} \, \exp\left(\frac{\eta^2}{2} \int \de^2 x_1 \de^2 x_2 \, \phi(x_1) \, \phi(x_2)  \,  G_{\chi}(x_1, x_2) \right) \ ,
    \label{IF_double_scalar}
\end{equation}
where $G_{\chi}(x_1, x_2)$ is the environment two-point function. Then, the system effective theory is
\begin{equation}
	Z = Z_{\chi} \int [\cD \phi] \, \exp\left[- \int \de^2 x \left( \frac{1}{2} \, \partial_{\mu} \phi \, \partial^{\mu} \phi + \frac{1}{2} \, m^2 \,  \phi^2 \right) + \frac{\eta^2}{2} \int \de^2 x_1 \de^2 x_2 \, \phi(x_1) \, \phi(x_2)  \,  G_{\chi}(x_1, x_2) \right] \ .
\end{equation}

We now want to extend the Influence Functional to the replicated geometry using the two proposals above, and compute the entanglement entropy that would follow from each.
\subsubsection*{Entanglement entropy from scenario $A$}

Following proposal $A$, the Influence Functional reads
\begin{equation}
	\cF_{A, n} = Z_{\chi}^n \, \exp\left(\frac{\eta^2}{2} \sum_{i=1}^n \int_{\cM_i} \de^2 x_1 \, \de^2 x_2 \, \phi(x_1) \, \phi(x_2)  \,  G_{\chi}(x_1, x_2) \right) \ .
    \label{IF_double_scalar_replica_A}
\end{equation}
The factor $Z_{\chi}^n$ comes from the fact that the environment replicas do not talk to each other, as specified by \eqref{A_environment_BC}. For this reason we have $n$ times the environment action, which contributes in the partition function to $Z_{\chi}^n$. For the same reason, the non-local term in \eqref{IF_double_scalar} gets upgraded to the one in \eqref{IF_double_scalar_replica_A} as a sum over the different sheets $\cM_i$. The environment two point-function $G_{\chi}(x_1, x_2)$ is identical to the one present in \eqref{IF_double_scalar}, since it's independently defined for each sheet. On the other hand, the system's fields are continuous in the whole $\cM^n$, consistent with \eqref{EFT_matter_BC}. 

The effective theory for scenario $A$ is then 
\begin{multline}
	Z_{A, n} = Z_{\chi}^n \int [\cD \phi] \, \exp\left(-\int_{\cM^n} \de^2 x \left( \frac{1}{2} \, \partial_{\mu} \phi \, \partial^{\mu} \phi + \frac{1}{2} \, m^2 \,  \phi^2 \right) \right. \\
	\left. + \, \frac{\eta^2}{2} \sum_{i = 1}^n \int_{\cM_i} \de^2 x_1 \, \de^2 x_2 \, \phi(x_1) \, \phi(x_2)  \,  G_{\chi}(x_1, x_2) \right) \ .
    \label{matt_eff_action_A}
\end{multline}
Let us compute the entanglement entropy in a perturbative expansion in $\eta$, to the first non-trivial order. To do it we expand \eqref{matt_eff_action_A} perturbatively. The first non-trivial order is
\begin{equation}
	Z_{A, n} = Z_{\chi}^n  \, Z_{\phi, n} \left( 1 + \frac{\eta^2}{2} \sum_{i=1}^n \int_{\cM_i} \de^2 x_1 \, \de^2 x_2 \, G^{(n)}_\phi(x_1, x_2)  \,  G_\chi(x_1, x_2)  + \dots \right) \ .
	\label{effective_theory_replicated}
\end{equation}
In Appendix \ref{sec:n-sheeted_Greens_Functions} we find that
\begin{equation}
	\int_{\cM_i} \de^2 x_1 \, \de^2 x_2   \, G^{(n)}_\phi(x_1, x_2)  \,  G_\chi(x_1, x_2) = \frac{1}{12 m^4 n} \ ,
	\label{GnG_product}
\end{equation}
which implies that
\begin{equation}
	Z_{A, n} = Z_{\chi}^n  \, \,  Z_{\phi, n} \left( 1 + \frac{\eta^2}{24 m^4} + \dots \right) \ .
	\label{Z_n_eff}
\end{equation}
Notice that the term proportional to $\eta^2$ in the RHS of \eqref{Z_n_eff} is independent of $n$ because the factor of $n$ in the denominator in the RHS of \eqref{GnG_product} gets canceled by the fact that we are summing $n$ identical contributions in \eqref{effective_theory_replicated}. Finally, we can use \eqref{Z_n_eff} to compute the entanglement entropy of the system's degrees of freedom in scenario $A$. Using the standard replica trick technique, it reads
\begin{equation}
	S_{A} = S_{\phi} + \frac{\eta^2}{24 m^4} + \dots \ .
	\label{S_A_simple_ex}
\end{equation}
Let us analyze this result. First, the term $S_{\phi}$ comes from $Z_{\phi, n}$, and it's the entanglement entropy of the system's degrees of freedom in the uncoupled limit \eqref{Entropy_for_massive_scalar_theory}. Then, the entanglement entropy of the environment is missing since $Z_{\chi}^n$ doesn't contribute to the entanglement entropy. Last, the term proportional to $\eta^2$ is a shift due to the coupling, which differs from one of the global theory \eqref{Entropy_interpretation}. This is not surprising, since there is no reason that the two quantities should coincide.

\subsubsection*{Entanglement entropy from scenario $B$}

Following proposal $B$, the Influence Functional reads
\begin{equation}
	\cF_{B, n} = Z_{\chi, n} \, \exp\left(\frac{\eta^2}{2} \int_{\cM^n} \de^2 x_1 \, \de^2 x_2 \, \phi(x_1) \, \phi(x_2)  \,  G_{\chi}^{(n)}(x_1, x_2) \right) \ .
    \label{IF_double_scalar_replica_B}
\end{equation}
The factor $Z_{\chi, n}$ in this case arises because, due to \eqref{B_environment_BC}, different environment replicas are coupled. This differs from scenario $A$ since $Z_{\chi, n} \neq Z_{\chi}^n$. For the same reason, the non-local term in \eqref{IF_double_scalar} gets upgraded to the one in \eqref{IF_double_scalar_replica_B} as a theory defined over the whole $\cM^n$. The environment two point-function then becomes $G_{\chi}^{(n)}(x_1, x_2)$, different from $G_{\chi}(x_1, x_2)$ since the former is defined in $\cM^n$ while the latter in $\cM$.

The effective theory for scenario $B$ is then 
\begin{multline}
	Z_{B, n} = Z_{\chi, n} \int [\cD \phi] \, \exp\left(-\int_{\cM^n} \de^2 x \left( \frac{1}{2} \, \partial_{\mu} \phi \, \partial^{\mu} \phi + \frac{1}{2} \, m^2 \,  \phi^2 \right) \right. \\
	\left. + \, \frac{\eta^2}{2} \int_{\cM^n} \de^2 x_1 \, \de^2 x_2 \, \phi(x_1) \, \phi(x_2)  \,  G_{\chi}^{(n)}(x_1, x_2) \right) \ .
    \label{matt_eff_action_B}
\end{multline}
Let us compute the entanglement entropy in a perturbative expansion in $\eta$, to the first non-trivial order. Expanding the exponential we have
\begin{equation}
	Z_{B,n} = Z_{\chi, n} \, Z_{\phi, n} \left( 1 + \frac{\eta^2}{2} \int_{\cM_n} \de^2 x_1 \de^2 x_2 \, \, G_{\chi}^{(n)}(x_1, x_2) \,  G_{\phi}^{(n)}(x_1, x_2) + \dots \right)  \ .
	\label{pert_exp_ZnB_in_eta}
\end{equation}
In Appendix \ref{sec:n-sheeted_Greens_Functions} we find that
\begin{equation}
	\int_{\cM_n} \de^2 x_1 \, \de^2 x_2   \, G^{(n)}(x_1, x_2) \, G^{(n)}(x_1, x_2) = \frac{1}{12 m^4 n} \ ,
\end{equation}
when appropriately regulated, which implies
\begin{equation}
	Z_{B, n} = Z_{n, \chi} \, Z_{n, \phi} \, \left( 1 + \frac{\eta^2}{24 m^4 n} + \dots \right) \ .
\end{equation} 
Finally, using the replica trick, we find
\begin{equation}
	S_{B} = S_{\phi} + S_{\chi} + \frac{\eta^2}{12 m^4 } + \dots  \ .
	\label{S_B_simple_ex}
\end{equation}
Interestingly, this result coincides, up to second order, to the entanglement entropy of the global system \eqref{Entropy_interpretation}. This is not surprising, since in both theories the environment replicas are coupled. For this reason it is expected that the two entropies \eqref{Entropy_interpretation} and \eqref{S_B_simple_ex} agree to all orders. 

We have thus shown that the two proposals $A$ and $B$ lead to quantitatively different results for the entanglement entropy.

\subsubsection*{Which prescription to use, or: {\it to couple or not to couple replicas?}}

\vspace{0.3cm}

At a first sight, we could conclude that generically only Scenario $A$ computes the entanglement entropy of the reduced density matrix, but this is not always the case. In the next Section we will argue that the most appropriate prescription to compute the reduced entanglement entropy for matter fields coupled to gravity is Scenario $B$. This conclusion holds also for every system that intrinsically, either by definition or to obtain the EFT description, has to deal with a coarse graining or an averaging, as we show in details in Appendix \ref{app:Coarse_grain_EE}. In particular, we claim that prescription $A$ computes a quenched version of the system entanglement entropy, while prescription $B$ computes and annealed version of the system entanglement entropy.

Proving this claim formally for matter fields coupled to gravity is particularly complicated, since a systematic averaging procedure to obtain a geometric bulk is still not known (see \cite{Saad:2019lba, Belin:2023efa} for progress in this direction). However, we will show that only prescription $B$ is consistent with a unitary answer, and only an annealed average can be consistent with a unitary answer as well. This gives a strong indication in favour of our previous claim. Moreover, the Influence Functional approach is particularly convenient since it does not require an exact microscopic knowledge of the underlying theory, but works only at the level of the EFT.

It is also worth pointing out that the $A$ boundary conditions can explicitly break a global or gauge symmetry present in the global system. Therefore, it is possible that the vacuum of the global theory is in a phase ({\it e.g.} where a global theory is spontaneously broken) that is realised in the non-replicated system and in the replicated system with $B$ boundary conditions, but not in the one with $A$. This rich phenomenology of possibilities indicates that a more careful investigation is required.

\section{The Influence Functional for gravity} \label{sec:Hawking_rad_EE}

We now turn our attention to Effective Field Theories of gravity coupled to matter. The aim of this Section is to compute the Influence Functional for matter minimally coupled to gravity in AdS$_d$, and then to use it to find the entanglement entropies associated to scenarios $A$ and $B$.

We always work in the approximation of weakly coupled gravity, which results also in a weak coupling between the gravitational and the matter sector. This is implemented by a small Newton constant $G_N$, which is necessary for the consistency of our effective field theory approach. From an EFT point of view, these assumptions are enough to guarantee that the factorized Hilbert space
\begin{equation}
	\cH = \cH_{\rm grav} \otimes \cH_{\rm mat}
	\label{Hilbert_space_gravity_matter}
\end{equation}
is a complete set of states. To find these states, one starts by choosing a gravitational saddle on top of which quantum fluctuations propagate. Then, quantizing the matter fields and the gravitons over this (generically curved) background gives $\cH_{\rm mat}$ and $\cH_{\rm grav}$, respectively.

The fact that the Hilbert spaces factorizes justifies the same open system interpretation we have used in previous sections. In particular, we can consider the gravitational sector to be an environment for the matter sector, which we view as the system under consideration. The Influence Functional then implements the (non-local) effective dynamics for the matter degrees of freedom. This language is particularly useful when studying the black hole information paradox. Considering the global state $\rho$ of an evaporating black hole, prepared through an appropriate Euclidean path integral, one can compute the entanglement entropy of the matter degrees of freedom considering the reduced density matrix 
\begin{equation}
    	\rho_{\rm mat} = \op{Tr}_{\rm grav} \left[ \rho \right] \ ,
\end{equation}
and then computing 
\begin{equation}
	S_{\rm mat} = - \op{Tr}\left[ \rho_{\rm mat} \log \left( \rho_{\rm mat} \right) \right]  \ .
	\label{S_mat_gravitational_EFT}
\end{equation}
As we have seen in previous sections, this can be done systematically using the Influence Functional. Indeed, we can do it explicitly. 

As anticipated, our low-energy action consists of gravity in $d$-dimensional (asymptotically) Anti-de Sitter minimally coupled to matter. This can be written as 
\begin{equation}
	I[g, \phi] = \frac{2}{\kappa^2} \int_{\cM} \de^d x \, \sqrt{g} \, \Big( R[g] - \Lambda \Big) + \frac{1}{\kappa^2} \int_{\partial\cM} \de^{d-1} y \, \sqrt{\gamma} \, K[\gamma] + I_{c.t.} + \int_{\cM \, \cup \, R} \de^d x \, \sqrt{g} \, \cL_{\rm mat}[g, \phi] \ .
	\label{class_act_grav}
\end{equation}
where $\kappa^2 = 32 \pi G_N$. In the action \eqref{class_act_grav}, $\cM$ is the (asymptotically) AdS$_d$ spacetime where gravity is dynamical, which can be any saddlepoint of the Einstein equations. At the boundary of $\cM$ we can also couple a flat region $R$. In this region gravity is non-dynamical, and only the matter fields are defined there. Because of this, it acts as a reservoir where the Hawking radiation is collected. We have included it for sake of generality. All the other terms in \eqref{class_act_grav} are the standard Einstein-Hilbert term with a cosmological constant, the Gibbons-Hawking-York term, the ensemble of counter-terms needed to regulate the theory, and the action for the matter fields. As anticipated, the latter are minimally coupled to gravity, even though this will not play a major role in the discussion. We believe that all the ideas presented here can be generalized for the case of non-minimal coupling.

The effective theory can be obtained expanding in perturbations $h_{\mu \nu}$ of the saddle $\hat g_{\mu \nu}$, the background metric of $\cM$. Precisely, we expand in
\begin{equation}
	g_{\mu \nu} = \hat g_{\mu \nu} + \kappa \, h_{\mu \nu} \ ,
\end{equation}
so that 
\begin{multline}
	I[h, \phi] = \frac{2}{\kappa^2} \int_{\cM} \de^d x \, \sqrt{\hat g} \, \Big( R[\hat g] - \Lambda \Big) + \frac{1}{\kappa^2} \int_{\partial \cM} \de^{d-1} y \, \sqrt{\hat \gamma} \, K[\hat \gamma] + I_{c.t.}[\hat g]\\ 
	+  \int_{\cM} \de^d x \, \sqrt{\hat g} \, \cL_{\rm g}[h] + \int_{\cM \cup R} \de^d x \, \sqrt{\hat g} \, \cL_{\rm mat}[\hat g, \phi] + \kappa \int_{\cM} \de^d x \, \sqrt{\hat g} \, h_{\mu \nu} \, T^{\mu \nu}[\hat g, \phi] + \dots  \ .
	\label{action_pert_exp}
\end{multline}
The first three terms in the RHS of \eqref{action_pert_exp} come from the gravitational action being evaluated on the saddle. The fourth term comes from the perturbative expansion of the gravitational action, and contains the kinetic term for the graviton, as well as all the interactions between them. The fifth term describes the matter effective theory on the fixed (generically curved) background of the saddle. Finally, we have an infinite series, weighted by powers of $\kappa$, describing the interactions between gravitons and matter fields. The sixth term in the RHS of \eqref{action_pert_exp} is the first non-trivial order of this series, where $T^{\mu \nu}[\hat g, \phi]$ is the stress energy tensor of the matter theory on the classical saddle.

The Influence Functional is then, by definition, 
\begin{multline}
	\cF[\phi] =  \exp \left(- \frac{2}{\kappa^2} \int_{\cM} \de^d x \, \sqrt{\hat g} \, \Big( R[\hat g] - \Lambda \Big) - \frac{1}{\kappa^2} \int_{\partial \cM} \de^{d-1} y \, \sqrt{\hat \gamma} \, K[\hat \gamma] - I_{c.t.}[\hat g]  \right) \\
	\times \int [\cD h ] \, \exp \left( -  \int_{\cM} \de^d x \, \sqrt{\hat g} \, \cL_{\rm g}[h] - \kappa \int_{\cM} \de^d x \, \sqrt{\hat g} \, h_{\mu \nu} \, T^{\mu \nu}[\hat g, \phi] + \dots \right) \ .
 \label{Def_IF_EFT_grav}
\end{multline}
The first term in the RHS of \eqref{Def_IF_EFT_grav} is (the exponential of) the on-shell gravitational action. For brevity we define
\begin{equation}
    \frac{2}{\kappa^2} \int_{\cM} \de^d x \, \sqrt{\hat g} \, \Big( R[\hat g] - \Lambda \Big) + \frac{1}{\kappa^2} \int_{\partial \cM} \de^{d-1} y \, \sqrt{\hat \gamma} \, K[\hat \gamma] + I_{c.t.}[\hat g] \equiv I_{g}[\hat g] \ .
\end{equation}
On the other hand, for the metric fluctuation we have to resort to a perturbative expansion in $\kappa$. The first non-trivial term is
\begin{multline}
    \int [\cD h ] \, \exp \left( -  \int_{\cM} \de^d x \, \sqrt{\hat g} \, \cL_{\rm g}[h] - \kappa \int_{\cM} \de^d x \, \sqrt{\hat g} \, h_{\mu \nu} \, T^{\mu \nu}[\hat g, \phi] + \dots \right) \\
    = \, Z_h[\hat g] \, \exp \left( \frac{\kappa^2}{2} \int_{\cM} \de^d x \, \sqrt{\hat g(x)} \, \int_{\cM} \de^d y \, \sqrt{\hat g(y)} \, G_{\mu \nu,\rho \sigma}(x, y) \, T^{\mu \nu}(x) \, T^{\rho \sigma}(y) + \dots \right) \ ,
    \label{metric_fluctuations_path_integral}
\end{multline}
where
\begin{equation}
    Z_h[\hat g] = \int [\cD h ] \, \exp \left( -  \int_{\cM} \de^d x \, \sqrt{\hat g} \, \cL_{\rm g}[h] \right)
\end{equation}
and 
\begin{equation}
    G_{\mu \nu,\rho \sigma}(x, y) = \frac{1}{Z_h[\hat g]}\int [\cD h ] \, h_{\mu \nu}(x)\, h_{\rho \sigma}(y) \, \exp \left( -  \int_{\cM} \de^d x \, \sqrt{\hat g} \, \cL_{\rm g}[h] \right) 
\end{equation}
is the graviton two-point correlation function. Notice that in the RHS of \eqref{metric_fluctuations_path_integral} we have already re-summed the series, so that in the graviton two-point correlation function we should only consider connected contributions. The Influence Functional is then
\begin{equation}
	\cF[\phi] =  e^{- I_g[\hat g]}\, Z_h[\hat g] \, \exp \left(\frac{\kappa^2}{2} \int_{\cM} \de^d x \,\de^d y \, \sqrt{\hat g(x) \hat g(y)} \,  G_{\mu \nu,\rho \sigma}(x, y) \, T^{\mu \nu}(x) \, T^{\rho \sigma}(y) + \dots \right)\ ,
 \label{IF_EFT_grav}
\end{equation}
and the effective theory for the matter sector is
\begin{multline}
	Z  =  \int [\cD \phi] \, \cF[\phi] \, \exp \left( - \! \int_{\cM} \de^d x \, \sqrt{\hat g} \, \cL_{\rm mat}[\hat g, \phi] \right) =  e^{- I_g[\hat g]}\, Z_h[\hat g] \int [\cD \phi] \, \exp \left( - \! \int_{\cM} \de^d x \, \sqrt{\hat g} \, \cL_{\rm mat}[\hat g, \phi] \right. \\
    \left. + \, \frac{\kappa^2}{2} \int_{\cM} \de^d x \,\de^d y \, \sqrt{\hat g(x) \hat g(y)} \, G_{\mu \nu,\rho \sigma}(x, y) \, T^{\mu \nu}(x) \, T^{\rho \sigma}(y) + \dots \right) \ .
    \label{Effective_Theory_matter_sector}
\end{multline}
This is the effective theory for the matter degrees of freedom in the presence of dynamical gravity, at first non-trivial order in $\kappa$. The Influence Functional contributes with a term which, superficially, resembles a $T \overline T$ deformation in higher dimensions, but would appear to lack the irrelevant RG flow interpretation of the latter.

This result can be used as-it-is to compute correlation functions of matter operators. As pointed out before, the result will be identical to the one obtained from the total action \eqref{action_pert_exp}. However, we will dedicate the next Section to use it to compute the matter entanglement entropy.

\subsection{Scenario $A$ - Hawking's result}

Following the idea of Section \ref{sec:warm_up}, we approach the calculation of the matter entanglement entropy using the replica trick. The replica manifold is fixed by the boundary conditions. It depends on which specific entanglement entropy one wishes to compute, which in turns determines how the manifold is replicated. As before, we call the replicated manifold $\cM^n$, and we think of it as a union of $n$ sheets $\cM_i$, identically to \eqref{Union_of_sheets}.

Now we can extend the gravitational Influence Functional following Scenario $A$. As before, the only interesting boundary condition for the matter fields is
\begin{equation}
    \phi(\cB_i^-) = \phi(\cB_{i+1}^+) \ ,
\end{equation}
similarly to \eqref{EFT_matter_BC}. On the other hand, gravitons of different sheets do not communicate. Each sheet has its own graviton field $h_{\mu \nu}^i$, and the boundary condition can be expressed as
\begin{equation}
    h^i_{\mu \nu}(\cB_i^-) =  h^i_{\mu \nu}(\cB_{i}^+) \ . 
    \label{A_graviton_BC}
\end{equation}
This can also be translated in the language of two point correlation functions $G_{\mu \nu,\rho \sigma}(x, y)$. In this case we do not have to change anything. Such correlation functions are naturally defined on each sheet $\cM_i$, and the complete Influence Functional reads
\begin{multline}
	\cF_{n}^{A}[\phi] =  e^{- n I_g[\hat g]}\, Z^n_h[\hat g] \, \, \times \\
\exp \left(\frac{\kappa^2}{2} \sum_{i = 1}^n \int_{\cM_i} \de^d x \,\de^d y \, \sqrt{\hat g(x) \hat g(y)} \,  G_{\mu \nu,\rho \sigma}(x, y) \, T^{\mu \nu}(x) \, T^{\rho \sigma}(y) + \dots \right)\ .
    \label{IF_EFT_grav_A}
\end{multline}
The first term in the RHS comes from the fact that $I_g[\hat g^n] = n I_g[\hat g]$. Notice how this implies that we have {\it not} included in $I_g[\hat g^n]$ any contribution from the conical singularities of $\cM^n$. A similar explanation can be given for the second term in the RHS, since $Z_h[\hat g^n] = Z^n_h[\hat g]$ due to the boundary conditions \eqref{A_graviton_BC}. Last, the $T^2$ term is just the sum over the single terms for each sheet.

We then use this proposal to compute the matter entanglement entropy, obtaining 
\begin{equation}
    S_{A} = S_{\phi}[\hat g] + \cO(\kappa^2)  \ . \label{S_A}
\end{equation}
This result includes only the entanglement entropy of matter fields in the static background $\hat g$. This term comes from the corresponding term in \eqref{Effective_Theory_matter_sector}, dressed with the boundary conditions \eqref{EFT_matter_BC}. The term $e^{- n I_g[\hat g]}$ does not give any contribution to the matter entanglement entropy $ S_{A}$, since it is canceled by an identical term in $Z^n$. Finally, an infinite series of corrections is present due to the graviton-matter exchanges, all proportional to the powers of the coupling $\kappa$, starting at order $\kappa^2$ from the $T^2$ term.

At leading order, the entropy $S_A$ is Hawking's result, namely the entropy of matter fields on a fixed curved background. Thus, it is natural to associate the $A$ boundary conditions to Hawking's calculation. Then, the procedure described can systematically compute gravitational corrections to Hawking result to all orders, in principle. Famously, for an evaporating black hole, this entanglement entropy is not consistent with a global unitary evolution. The corrections proportional to powers of $\kappa$ are believed to be small throughout most of the evolution, since the gravitational EFT is a good approximation until the BH approaches the Planck length. Much earlier in the evolution, around the so-called {\it Page time}, it is possible to find macroscopic deviations from a global unitary evolution.  

The proposed solution \cite{Almheiri:2019qdq, Penington:2019kki} finds indeed new dominant replica saddle-points, which in our language appear considering Scenario $B$. We dedicate the next Section to exploring this issue.

\subsection{Scenario $B$ - Quantum Extremal Surfaces}

In recent years it has been shown how, for gravitational theories in Anti de-Sitter coupled to a reservoir region $R$, the entanglement entropy
\begin{equation}
    S(R) = \operatorname{min}_{I} \left\{ \frac{8 \pi A(\partial I)}{\kappa^2} + S_{\rm bulk \, \, fields} (I \cup R)\right\}
    \label{Island_Formula}
\end{equation}
is consistent with a unitary evolution \cite{Penington:2019npb, Almheiri:2019psf}. This is the so-called {\it island formula}, a generalization of the Ryu-Takayanagi proposal \cite{Ryu:2006ef} to compute entanglement entropies in holographic theories which includes quantum corrections. The prescriptions entails minimizing the functional in the RHS of \eqref{Island_Formula}, considering all regions $I$ in the bulk. The boundary $\partial I$ is called {\it Quantum extremal surface}, while the interior region $I$ is the {\it island}. This formula is quite different from the previous ones found \eqref{S_A}. In particular, the two key ingredients that make the evolution consistent with unitarity are the area term and the minimization present in \eqref{Island_Formula}. 

In a series of works employing the replica trick it was shown how this formula can be justified considering smooth saddles of the replicated geometry, either as a way to compute entanglement entropies in holographic theories \cite{Lewkowycz:2013nqa, Faulkner:2013ana, Dong:2017xht} to just semiclassical gravity coupled to a bath \cite{Penington:2019kki, Almheiri:2019qdq}. In particular, these smooth replica saddles are responsible for both the area term and the minimization of the island formula \eqref{Island_Formula}. The resulting entanglement entropy \eqref{Island_Formula} can differ by a non-perturbative quantity with respect to the result \eqref{S_A} obtained in the previous Section.

We now want to argue that the matter entanglement entropy computed from Scenario $B$ leads to the island formula \eqref{Island_Formula}. In this context we will find new saddlepoint solutions in the replicated geometry, which are needed to restore unitarity. Therefore, the entanglement entropy for matter fields on a gravitational background using Scenario $B$ can be obtained following these steps.

\begin{itemize}
    \item Replicate the global system (matter and gravity) as specified by the boundary conditions of interest. 
    \item Find all smooth saddlepoint geometries $\cM_n$ with the same boundary conditions. We call their metric $\hat g_n$. Notice the difference with $\cM^n$ and $\hat g^n$ of the previous section, which had conical singularities. Here we are considering only smooth saddlepoints. 
    \item Consider the effective theory on these backgrounds, composed by metric and matter fluctuations. Then integrate out the metric fluctuations to obtain the matter effective theory on this geometry.
\end{itemize}

\begin{figure}[t]
    \centering
    \includegraphics[width=0.8\textwidth]{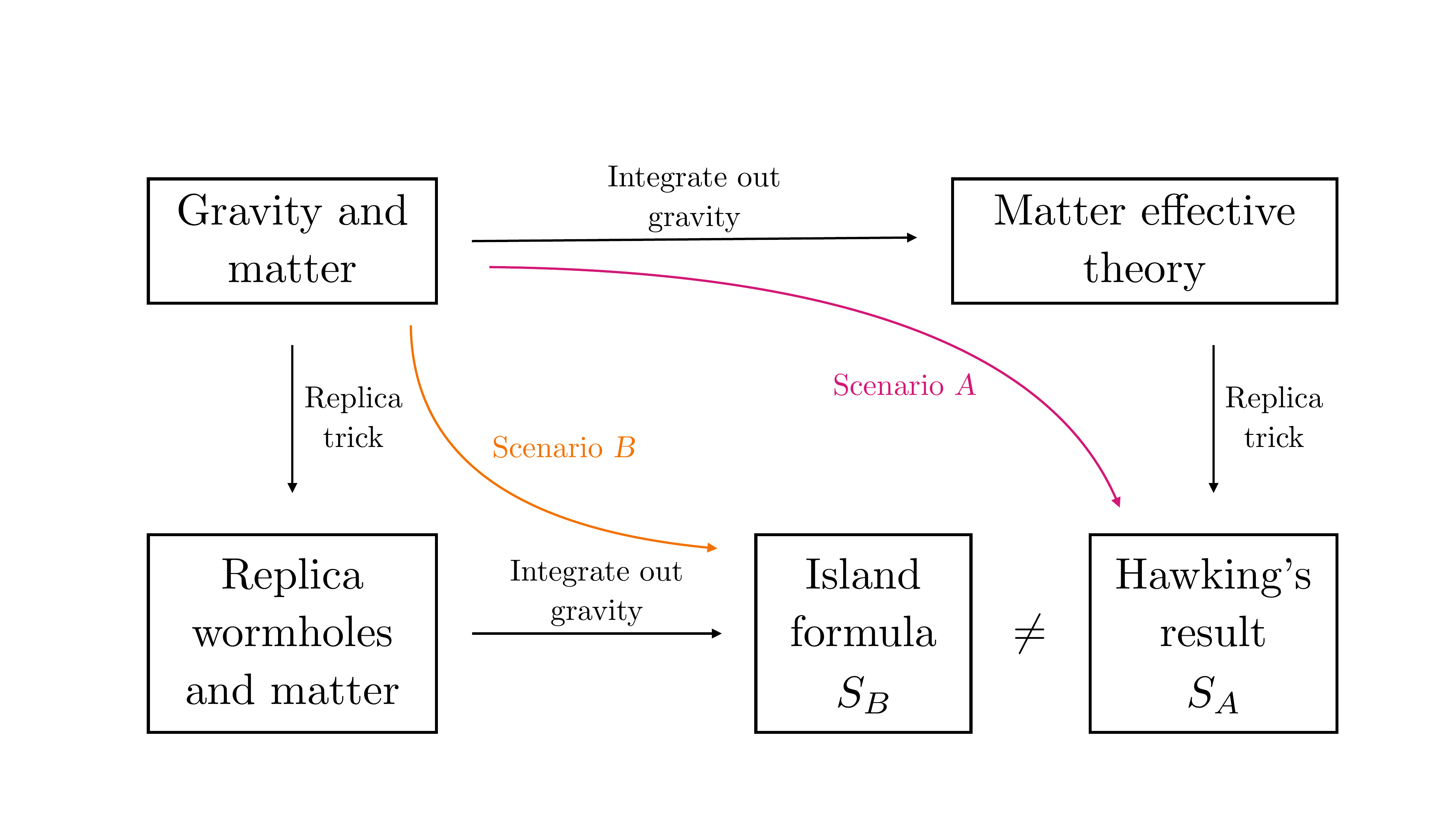}
    \caption{Non-commutative diagram representing the differences between Scenario $A$ to compute the replicated Open EFT (and then the reduced matter entanglement entropy), and Scenario $B$, which seems to be better suited for quantum gravity.}
    \label{non_commutative_diagram}
\end{figure}

The key difference with the approach of the previous Sections is that here one first has to replicate, and then compute the Influence Functional integrating out metric fluctuations of the smooth replicated saddle. In particular, considering smooth gravitational replica saddles instead of ones with conical defects is a key ingredient. We summarize this points in Figure \ref{non_commutative_diagram}. Notice moreover that this difference arises only for the replicated theory, and for this reason the Open EFT \eqref{Effective_Theory_matter_sector} computed previously can still be used for any observable not involving the replica trick.

We now compute quantitatively the Open EFT for this prescription and show that the resulting matter entanglement entropy is the island formula. The Influence Functional in this case is
 \begin{multline}
	\cF_n^B[\phi] =  e^{- I_g[\hat g_n]}\, Z_h[\hat g_n] \, \, \times \\
    \exp \left(\frac{\kappa^2}{2} \int_{\cM_n} \de^d x \,\de^d y \, \sqrt{\hat g_n(x) \hat g_n(y)} \,  G_{\mu \nu,\rho \sigma}(x, y) \, T^{\mu \nu}(x) \, T^{\rho \sigma}(y) + \dots \right)\ ,
    \label{IF_grav}
\end{multline}
where now the graviton correlation function is defined in the smooth replicated saddle, thus
\begin{equation}
    G_{\mu \nu,\rho \sigma}(x, y) = \frac{1}{Z_h[\hat g_n]}\int [\cD h ] \, h_{\mu \nu}(x)\, h_{\rho \sigma}(y) \, \exp \left( -  \int_{\cM_n} \de^d x \, \sqrt{\hat g_n} \, \cL_{\rm g}[h] \right) \ .
\end{equation}
Then, the replicated matter Open EFT is in this prescription
\begin{equation}
    Z_n = \int [\cD \phi] \, \cF_n^B [\phi] \, \exp \left( - \! \int_{\cM_n} \de^d x \, \sqrt{\hat g_n} \, \cL_{\rm mat}[\hat g_n, \phi] \right) \ .
\end{equation}

The expression for the replicated Influence Functional \eqref{IF_grav} can also be simplified. To do it we use the idea of {\it replica cosmic branes} \cite{Dong:2016fnf}. Let us consider the orbifolded space
\begin{equation}
    \tilde \cM_n = \cM_n / \mathbb Z_n \ . 
\end{equation}
Based on what kind of boundary condition we choose for the replica geometry, $\tilde \cM_n$ may or may not have fixed points of the orbifold. If it does, these are generally codimension-2 manifolds which are called replica cosmic branes. Using the fact that $\cM_n$ is smooth, at the location of these branes $\tilde \cM_n$ has conical singularities of opening angles 
\begin{equation}
    \Delta \theta = 2 \pi \left(1 - \frac{1}{n} \right)  \ .
    \label{Conical_defects_branes}
\end{equation}
Effectively, we can take them into account introducing the corresponding term in the action, namely
\begin{equation}
    I[\tilde \cM_n] = n \, I[\cM_n] + T_n \int_{\tilde \cM_n} \de^{d-2} y \, \sqrt{\hat \gamma_n} \ ,
    \label{Orbifolded_action_with_branes}
\end{equation}
where the value of their tension, 
\begin{equation}
    T_n = \frac{1}{4G_N} \left(1 - \frac{1}{n}\right)
\end{equation}
is obtained from the Einstein equations of motion for a conical defect of magnitude \eqref{Conical_defects_branes}. It is important to stress that the action \eqref{Orbifolded_action_with_branes} should only be used on shell to determine the position of such cosmic branes, but their contribution should not be taken into account in any other way, since the original action $ I[\cM_n]$ doesn't contain any singularity.

The orbifolded action can then be used to simplify the expression for the Influence Functional. To do it we follow \cite{Lewkowycz:2013nqa, Faulkner:2013ana} and consider its variation with respect to the number of replicas $n$. The key point is that on-shell, the variation of the gravitational action is a boundary term. Close to the conformal boundary it vanishes, while close to the cosmic branes, where we regulate introducing a cylinder of radius $r = \ve$ around them, it doesn't. To be precise one has
\begin{multline}
    \partial_n \log \tilde Z_n = - \int E_g \partial_n \tilde g_n - \int E_\phi \partial_n \tilde g_n \\ 
    - \frac{2}{\kappa^2} \int_{r = \ve} \! \! \! \de^{d-1} y \, \sqrt{\tilde \gamma_n} \, \big( \nabla^\mu \partial_n g_{\mu r} - g^{\mu \nu} \nabla_r \partial_n g_{\mu \nu}\big) + \partial_n \big( \log Z_h[\tilde g_n] + \log Z_\phi[\tilde g_n] + \dots \big) \ .
    \label{Variation_orbifolded_action}
\end{multline}
The first two terms in the RHS are the Einstein equations and they jointly vanish, since we have already imposed them to find $\tilde g_n $. The third term is the boundary term that we evaluate close to the cosmic branes. One can approximate the metric there as
\begin{equation}
    \de s^2 = \de r^2 + \frac{r^2}{n^2} \, \de \theta^2 + \gamma_{ij}(y) \de y^i \de y^j \ .
\end{equation}
This implements the correct conical singularity provided $\theta$ is $2 \pi$ periodic. We can evaluate the third term in the RHS of \eqref{Variation_orbifolded_action} with this metric to find
\begin{equation}
    \frac{2}{\kappa^2} \int_{r = \ve} \! \! \! \de^{d-1} y \, \sqrt{\tilde \gamma_n} \, \big( \nabla^\mu \partial_n g_{\mu r} - g^{\mu \nu} \nabla_r \partial_n g_{\mu \nu}\big) = \frac{8 \pi A[{\rm CB}_n]}{\kappa^2 n^2} \ ,
\end{equation}
where in the RHS we have defined the area of the $n$-cosmic brane. We can then integrate in $n$ to find $\log \tilde Z_n$. We can neglect any constant of integration, since it will not contribute to the dynamics and neither to the final entanglement entropy. Last, multiplying by $n$ we get 
\begin{equation}
    \log Z_n = \frac{8 \pi A[{\rm CB}_n]}{\kappa^2} + \log Z_h[\hat g_n] + \log Z_\phi[\hat g_n] + \dots \ ,
\end{equation}
or, in a more extended form
\begin{multline}
	Z_n  =  \exp \left(\frac{8 \pi A[{\rm CB}_n]}{\kappa^2} \right) \, Z_h[\hat g_n] \int [\cD \phi] \, \exp \left( - \! \int_{\cM_n} \de^d x \, \sqrt{\hat g_n} \, \cL_{\rm mat}[\hat g_n, \phi] \right. \\
    \left. + \, \frac{\kappa^2}{2} \int_{\cM_n} \de^d x \,\de^d y \, \sqrt{\hat g_n(x) \hat g_n(y)} \, G_{\mu \nu,\rho \sigma}(x, y) \, T^{\mu \nu}(x) \, T^{\rho \sigma}(y) + \dots \right) \ .
    \label{Z_n_gravity_simplified}
\end{multline}
Thus the Influence Functional reads
\begin{multline}
	\cF_n^B[\phi] =  \exp \left(\frac{8 \pi A[{\rm CB}_n]}{\kappa^2} \right)\, Z_h[\hat g_n] \, \, \times \\
    \exp \left(\frac{\kappa^2}{2} \int_{\cM_n} \de^d x \,\de^d y \, \sqrt{\hat g_n(x) \hat g_n(y)} \,  G_{\mu \nu,\rho \sigma}(x, y) \, T^{\mu \nu}(x) \, T^{\rho \sigma}(y) + \dots \right)\ .
    \label{IF_grav_simplified}
\end{multline}
Our last goal of this section is to compute the entanglement entropy associated to \eqref{Z_n_gravity_simplified} through the Influence Functional \eqref{IF_grav_simplified} and compare it to the island formula \eqref{Island_Formula}. The area term clearly gives a similar contribution in the entanglement entropy. Moreover, the entropy from the quantum fields also give corresponding contributions. However, we are still missing the extremization characteristic of the island formula. Indeed, while it is true that the position of the cosmic branes is determined by an extremization, the procedure is different. For the cosmic branes we have only used the on-shell equations of motions while for the island formula one takes into account also quantum corrections. Luckily, we can use the result of \cite{Faulkner:2013ana}, which found that the entanglement entropy associated to the quantum fluctuations also takes into account the corresponding change in area given by the change of the gravitational saddle due to quantum corrections. Then, following the result of \cite{Faulkner:2013ana}, the entanglement entropy of the matter sector from \eqref{IF_grav_simplified} reads
\begin{equation}
    S_B = \frac{8 \pi A[{\rm CB}_
    {n \to 1}]}{\kappa^2} + \frac{8 \pi \delta A}{\kappa^2} + S_h[\hat g] + S_\phi[\hat g] + \cO(\kappa^2)\ .
    \label{Island_exp}
\end{equation}
It can be shown \cite{Engelhardt:2014gca} that this procedure agrees with \eqref{Island_Formula} to all orders in $\kappa^2$. Interestingly, graviton fluctuations contribute to the matter entanglement entropy. This may seem a contradiction, since in the original formulation of the problem we wanted to integrate them out. On the other hand, \eqref{Island_exp} is consistent with unitarity due to the presence of the area of the cosmic branes. We believe that a prescription that does include the area of the cosmic branes but does not include the graviton fluctuations will not be as natural as Scenario $B$, and therefore we follow the latter. However, it would be interesting to have an independent reasoning to include graviton fluctuations in the matter entanglement entropy coupled to gravity.

All in all, we have then found a replicated Influence Functional \eqref{IF_grav_simplified} for the matter sector whose reduced entanglement entropy is consistent with unitarity, which then seems to be better suited to describe the gravitational Open EFT. 

We would also like to stress that, even if the prescription presented in this Section to compute the replicated Influence Functional is motivated by the island formula, we believe that the resulting Open EFT can potentially have deeper implications, being a more general framework. We will try to address some of them in Section \ref{sec:Discussion}.

\section{An application: JT gravity and the island model} \label{JT_gravity_and_isl_model}

In this Section we want to show a concrete example to apply the ideas just presented. We are going to focus on a two-dimensional model where JT gravity\footnote{For a nice review on JT gravity and solvable models of quantum black holes see \cite{Mertens:2022irh}.} is coupled to two-dimensional CFT which extends also in a non-gravitating reservoir, first proposed in \cite{Almheiri:2019psf} (see also \cite{Almheiri:2019qdq}). On a technical level, the analysis performed in Section \ref{sec:Hawking_rad_EE} assumed at least four-dimensional gravity, since we allowed metric fluctuations. On the other hand, low-dimensional models are convenient setups since they allow analytical solutions, and we will see that the physics behind them doesn't differ much from their higher dimensional counterparts.

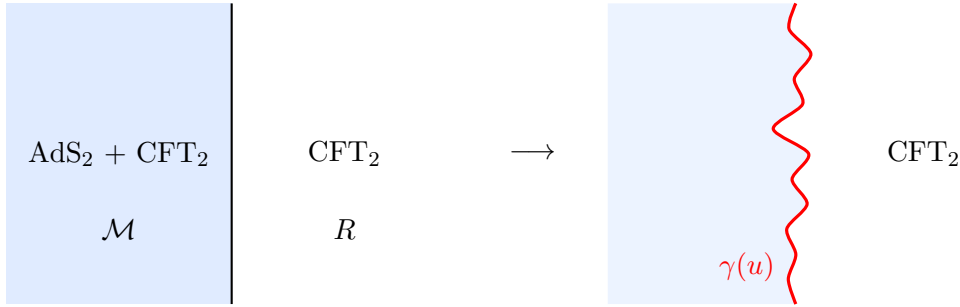
\begin{figure}
\centering
\begin{tikzpicture}
\fill[gravity]  (0,0) -- (0,4) -- (-3,4) -- (-3,0) -- cycle;
\draw[thick] (0,0) -- (0,4);

\draw[black] (-1.5,2) node{AdS$_2$ + CFT$_2$};

\draw[black] (-1.5,1) node{$\cM$};

\draw[black] (1.5,2) node{CFT$_2$};

\draw[black] (1.5,1) node{$R$};

\draw[black] (4,2) node{$\longrightarrow$};

\fill[gravity, opacity = 0.6] plot [smooth, tension=0.5] coordinates { (7.5,0) (7.4,0.33) (7.5,0.66) (7.38,1) (7.65,1.33) (7.45,1.66) (7.67,2) (7.2,2.33) (7.59,2.66) (7.48,3) (7.7,3.33) (7.41,3.66) (7.5,4) } -- (5,4) -- (5,0) -- cycle;

\draw[very thick, red] plot [smooth, tension=0.5] coordinates { (7.5,0) (7.4,0.33) (7.5,0.66) (7.38,1) (7.65,1.33) (7.45,1.66) (7.67,2) (7.2,2.33) (7.59,2.66) (7.48,3) (7.7,3.33) (7.41,3.66) (7.5,4) };

\draw[black] (9.2,2) node{CFT$_2$};

\draw[] (0, -0.5) node{};
\draw[red] (6.85, 0.5) node{$\gamma(u)$};

\end{tikzpicture}
\caption{JT gravity coupled to a two-dimensional CFT. The gravitational theory is defined on a saddle $\cM$, while the CFT$_2$ on $\cM \cup R$, where $R$ is a non-gravitational reservoir. The gravitational dynamics gest localized on the AdS$_2$ boundary through curves $\gamma(u)$ of length is $\beta / \ve$, so that the renormalized length of the boundary is $\beta$.}
\label{fig:2D_BH}
\end{figure}

\begin{figure}[t]
\begin{center}
\begin{tikzpicture}

\draw[very thick, gray] plot [smooth, tension=1] coordinates {(2.5,0) (3.1,0.25) (3.5,0.9) };
\draw[very thick, gray] plot [smooth, tension=1] coordinates {(2.5,2) (3.1,1.75) (3.5,1.1) };

\draw[very thick, gray] (0,0) -- (2.5,0);
\draw[very thick, gray] (0,2) -- (2.5,2);
\draw[very thick, CadetBlue, name path=D] (0,0) -- (-1,0);
\draw[very thick, CadetBlue, name path=E] (0,2) -- (-1,2);

\draw[very thick, CadetBlue, dashed] (0,2) arc (90:270:0.5 and 1);

\draw[very thick, CadetBlue] plot [smooth, tension=1] coordinates {(-1,0) (-1.6,0.25) (-2,0.9) };
\draw[very thick, CadetBlue] plot [smooth, tension=1] coordinates {(-2,1.1) (-1.6,1.75) (-1,2)};

\fill[gravity, opacity = 0.5]  (0,0) arc (-90:90:0.5 and 1) -- (0,2) -- (-1,2) -- plot [smooth, tension=1] coordinates {(-1,2) (-1.6,1.75) (-2,1.1) } -- plot [smooth, tension=1] coordinates {(-2,0.9) (-1.6,0.25) (-1,0)} -- cycle;
\fill[gravity, opacity = 0.5]  (0,0) arc (270:90:0.5 and 1) -- (0,2) -- (-1,2) -- plot [smooth, tension=1] coordinates {(-1,2) (-1.6,1.75) (-2,1.1) } -- plot [smooth, tension=1] coordinates {(-2,0.9) (-1.6,0.25) (-1,0)} -- cycle;

\filldraw[red] (1.25,1) circle (2pt);

\draw[Green, very thick] (1.25,1) circle (0.25);

\draw [Green, very thick, decorate, decoration={snake, amplitude=.4mm, segment length=2mm, post length=0}] (1.5,1) -- (3.5,1);

\draw[very thick, CadetBlue] (0,0) arc (-90:90:0.5 and 1);

\end{tikzpicture}
\end{center}
\caption{The replicated geometry considered in this Section.}
\label{fig:b_definition}
\end{figure}
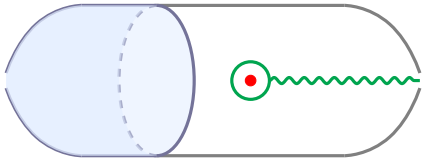

Concretely, we will compute the Influence Functional for a semi-infinite interval starting from a point in the radiation region, at $y = b$ in the coordinates of \eqref{Poinc-thermal}, to infinity, as shown in Figure \ref{fig:b_definition}. We find the Influence Functional using Scenario $A$ at leading order to be 
\begin{equation}
    \cF_{n}^A = \exp[n S_0 + \frac{8 \pi^2 \Phi_b n }{\kappa^2 \beta}] \ ,
    \label{Fn_naive}
\end{equation}
while the gravitational replicated Influence Functional at leading order to be 
\begin{equation}
    \cF_{n}^B = \exp[S_0 + \frac{8 \pi^2 \Phi_b }{\kappa^2 n \beta} \, \coth(\frac{2 \pi b}{\beta}) ] \ .
    \label{Fn_gravitational}
\end{equation}
Clearly, the former Influence Functional does not contribute to the entanglement entropy, while the gravitational one does, in a way consistent with the island formula. We will be able to show it quantitatively.

In the rest of the Section we will go through the technical details needed to arrive at \eqref{Fn_naive} and \eqref{Fn_gravitational}. As anticipated in general in Section \ref{sec:Hawking_rad_EE}, this will involve finding the replica wormhole smooth solutions connected to the replicated geometry of Figure \ref{fig:b_definition}. this will be achieved considering the conformal transformation that maps the asymptotic AdS$_2$ boundary into itself while mapping $y = b$ to infinity. We will argue that in these new coordinates the replica wormhole solution has a very simple form.

The action of interest for us is
\begin{equation}
	I = - \frac{S_0}{4 \pi} \int_{\cM} R - \frac{S_0}{2\pi} \int_{\partial \cM} K - \frac{2}{\kappa^2} \int_{\cM} \Phi (R + 2) - \frac{4}{\kappa^2} \int_{\partial \cM} \, \Phi (K-1) + I_{\rm CFT}(g, \phi) \ ,
\end{equation}
where we have chosen the AdS$_2$ radius to be unity. The matter action is the last term in the RHS, it is defined on $\cM \cup R$, and depends on the matter fields $\phi$ as well as the spacetime metric $g_{\mu \nu}$. As anticipated, we assume this matter theory to be a two-dimensional CFT with large central charge $c$. The other terms in the RHS are purely gravitational. The first two terms are topological, and combined they give the Euler characteristic of spacetime, 
\begin{equation}
    \frac{1}{4 \pi} \int_{\cM} R - \frac{1}{2\pi} \int_{\partial \cM} K = \chi_{\cM} = 2 - 2 g - n \ ,
\end{equation}
where $g$ and $n$ are the genus and the boundaries of $\cM$. This term can them be thought of a parameter that weights the various gravitational saddles, ordering their contribution to the gravitational path integral, provided we take $S_0$ to be positive and large. In this case, saddles which have a lower Euler characteristic are dominant. In the following we will only consider disk geometries, which are the dominant ones in the case of a single boundary, and they have $\chi = 1$. The other two terms in the gravitational Lagrangian involve a dilaton field $\Phi$. Its equation of motion imposes the constraint 
\begin{equation}
	R = -2 \ ,
    \label{dilaton_constr}
\end{equation}
which also holds quantum mechanically, where we would path-integrate $\Phi$. Locally, any saddle satisfying \eqref{dilaton_constr} can be described by the metric \cite{Mertens:2022irh}
\begin{equation}
	\de s^2 = \frac{\de T^2 + \de Z^2}{Z^2} = \frac{4}{\big(Y + \bar Y \big)^2} \, \de Y \, \de \bar Y  \ , \qquad \qquad Y = Z + i T\ .
\end{equation}
On the other hand, varying the metric we obtain the equations describing the dynamics of the dilaton, which are
\begin{equation}
	\nabla_{\mu} \nabla_{\nu} \Phi - g_{\mu \nu} \nabla^2 \Phi + g_{\mu \nu} \Phi = 0\ ,
\end{equation}
generically solved by \cite{Mertens:2022irh}
\begin{equation}
    \Phi = \frac{\alpha + \eta \, T + \delta (T^2 + Z^2)}{Z} \ .
\end{equation}
Imposing $R = -2$, the whole dynamics is contained in the extrinsic curvature, thus in the boundary. To describe this dynamics, we cut out a curve $\gamma(u) = \big( T(u), Z(u) \big)$ close to the boundary, and impose 
\begin{equation}
    \sqrt{h} \, \Big|_{\gamma(u)} = \frac{1}{\ve} \ , \qquad \qquad \Phi \, \Big|_{\gamma(u)} = \frac{\Phi_b(u)}{\ve} \ .
\end{equation}
This is solved at leading order by $Z(u) = \ve T'(u)$, so that the normal vector is
\begin{equation}
    n_{\mu}(u) = \frac{1}{Z(u) \sqrt{Z'(u)^2 + T'(u)^2}} \, \Big( - Z' (u) , T'(u) \Big) \ ,
\end{equation}
and 
\begin{equation}
    K = g^{\mu \nu} \nabla_{\mu} n_{\nu} = 1 + \ve^2 \left[ \frac{T'''(u)}{T'(u)} - \frac{3}{2} \left(\frac{T''(u)}{T'(u)} \right)^2 \right] + \dots = 1 + \ve^2 \left\{ T(u) , u \right\} + \dots \ .
\end{equation}
We can then rewrite the gravitational action as a Schwarzian action, namely
\begin{equation}
	I =  - \frac{4}{\kappa^2} \int \de u \, \, \Phi_b (u) \left\{ T(u) , u \right\} + I_{\rm CFT}(g, \phi) \ .
\end{equation}
This is a theory that can be thought of as living only on the boundary. To discuss black-hole physics, we transform the state from vacuum to the thermal state. This can be achieved considering the conformal transformation that maps the conformal vacuum to the thermal strip, which is
\begin{equation}
    Y = \tanh(\frac{\pi y}{\beta}) \ , \qquad \qquad y = \sigma + i \tau \ ,
    \label{Poinc-thermal}
\end{equation}
where now $\tau$ is $\beta$ periodic, namely $\tau \simeq \tau + \beta$. In these coordinates, the metric for the gravitational region $\cM$ is
\begin{equation}
    \de s^2_{\cM} = \frac{4\pi^2}{\beta^2} \frac{\de y \, \de \bar y}{\sinh[2](\frac{\pi}{\beta}(y + \bar y))}  \ .
    \label{metric_JT_CFT}
\end{equation}
The asymptotic boundary is at $\sigma = 0$, and we decide to place the gravitational region for $\sigma$ negative. In order to glue to it a flat radiation region, we cut off the asymptotically AdS$_2$ space at $\sigma = - \ve$ (where $\ve$ is positive), and we consider a flat space for $\sigma > - \ve$ such that the metric is continuous. All in all the metric is 
\begin{equation}
    \de s^2 = \begin{cases}
        \displaystyle \frac{4\pi^2}{\beta^2} \frac{\de y \, \de \bar y}{\sinh[2](\frac{\pi}{\beta}(y + \bar y))} \qquad & \text{for } \sigma < - \ve \qquad (\cM)\\
        \displaystyle \frac{\de y \, \de \bar y}{\ve^2} \qquad  & \text{for } \sigma > - \ve \qquad \, \,  (R)
        \end{cases} 
    \label{Original_geometry}
\end{equation}
This is the same convention used in \cite{Almheiri:2019qdq}. We only consider static solutions, and thus we take the dilaton to be, up to SL(2,$\mathbb R$) transformations, 
\begin{equation}
    \Phi = - \frac{2 \pi \Phi_b}{\beta} \coth(\frac{\pi}{\beta} \, (y + \bar y)) \ .
    \label{classical_thermal_dilaton}
\end{equation}
From \eqref{Poinc-thermal}, we also have the identification  
\begin{equation}
    T(u) = \tan(\frac{\pi \tau(u)}{\beta}) \ ,
\end{equation}
where $u$ now is a compact coordinate such that $\tau(u + \beta) = \tau + \beta$. This means that $\tau(u)$ is a diffeomorphism mapping the thermal circle back to the itself. The action is then 
\begin{equation}
	I =  - \frac{4\Phi_b}{\kappa^2} \int_0^\beta \de u \, \left\{ \tan(\frac{\pi \tau(u)}{\beta}) , u \right\} + I_{\rm CFT}(g, \phi) \ .
\end{equation}
and the partition function is
\begin{equation}
    Z = e^{S_0} \int_{{\rm Diff}(S^1)/{\rm SL}(2,\mathbb R)} [\cD \tau]  \, \exp(\frac{4\Phi_b}{\kappa^2} \int_0^\beta \de u \, \left\{ \tan(\frac{\pi \tau(u)}{\beta}) , u \right\} ) \int [\cD \phi] \, e^{-I_{\rm CFT}(g, \phi)}
\end{equation}
To be precise, also the matter theory depends on $\tau(u)$. In particular one would need to specify the boundary condition 
\begin{equation}
    \phi \Big|_{\gamma(u)} = \phi_b(u) \ ,
    \label{bc_matter_fields_cutouff_surface}
\end{equation}
and thus 
\begin{equation}
    \int [\cD \phi] \, e^{-I_{\rm CFT}(g, \phi)} = Z_{\rm mat} [\phi_b, \tau(u)] \ .
\end{equation}
In the semiclassical limit ($\Phi_b / \kappa^2 \gg 1$), we can evaluate the path integral by saddle-point. The equations of motion are
\begin{equation}
    \frac{\de}{\de u} \left\{ \tan(\frac{\pi \tau(u)}{\beta}) , u \right\} = 0 \ ,
    \label{non-replicated-EOM}
\end{equation}
which can be solved by
\begin{equation}
    \tau (u) = u \ , \qquad \qquad u \in [0, \beta) \ ,
    \label{cutoff_surface}
\end{equation}
and also leads to the relation $\sigma = - \ve$. We can then specify the matter boundary conditions \eqref{bc_matter_fields_cutouff_surface} on the cut-off surface \eqref{cutoff_surface}. Using the standard holographic dictionary in the semiclassical approximation, we can say that 
\begin{equation}
    Z_{\rm mat} [\phi_b, \tau(u)] = \int_{\phi|_{\gamma(u)} = \phi_b(u)} [\cD \phi] \, e^{-I_{\rm CFT}(g, \phi)} \ ,
\end{equation}
where the RHS is evaluated on the fixed background metric. This relation is valid up to corrections coming from back-reaction on the geometry due to the matter fields. Thus, to this level of approximation the Influence Functional does not explicitly depends on the boundary condition of matter fields, and it is
\begin{equation}
    \cF_{\rm JT} = e^{S_0} \int_{{\rm Diff}(S^1)/{\rm SL}(2,\mathbb R)} [\cD \tau]  \, \exp(\frac{4\Phi_b}{\kappa^2} \int_0^\beta \de u \, \left\{ \tan(\frac{\pi \tau(u)}{\beta}) , u \right\} ) \ .
\end{equation}
On the saddle \eqref{cutoff_surface}, the Schwarzian derivative is then 
\begin{equation}
    \left\{ \tan(\frac{\pi u}{\beta}) , u \right\} = \frac{2 \pi^2}{\beta^2} \ ,
\end{equation}
which implies
\begin{equation}
    \cF_{\rm JT} = \exp(S_0 + \frac{8 \pi^2 \Phi_b}{\kappa^2 \beta}) \ ,
\end{equation}
and that to leading order the Open EFT is 
\begin{equation}
    Z = \cF_{\rm JT} \int [\cD \phi] \, e^{-I_{\rm CFT}(g, \phi)} \ .
\end{equation}

\begin{figure}[t]
\begin{center}
\begin{tikzpicture}

\begin{scope}[shift={(0,0)}]

\draw[very thick, gray] plot [smooth, tension=1] coordinates {(2.5,0) (3.1,0.25) (3.5,0.9) };
\draw[very thick, gray] plot [smooth, tension=1] coordinates {(2.5,2) (3.1,1.75) (3.5,1.1) };

\draw[very thick, gray] (0,0) -- (2.5,0);
\draw[very thick, gray] (0,2) -- (2.5,2);
\draw[very thick, CadetBlue, name path=D] (0,0) -- (-1,0);
\draw[very thick, CadetBlue, name path=E] (0,2) -- (-1,2);

\draw[very thick, CadetBlue, dashed] (0,2) arc (90:270:0.5 and 1);

\draw[very thick, CadetBlue] plot [smooth, tension=1] coordinates {(-1,0) (-1.6,0.25) (-2,0.9) };
\draw[very thick, CadetBlue] plot [smooth, tension=1] coordinates {(-2,1.1) (-1.6,1.75) (-1,2)};

\fill[gravity, opacity = 0.5]  (0,0) arc (-90:90:0.5 and 1) -- (0,2) -- (-1,2) -- plot [smooth, tension=1] coordinates {(-1,2) (-1.6,1.75) (-2,1.1) } -- plot [smooth, tension=1] coordinates {(-2,0.9) (-1.6,0.25) (-1,0)} -- cycle;
\fill[gravity, opacity = 0.5]  (0,0) arc (270:90:0.5 and 1) -- (0,2) -- (-1,2) -- plot [smooth, tension=1] coordinates {(-1,2) (-1.6,1.75) (-2,1.1) } -- plot [smooth, tension=1] coordinates {(-2,0.9) (-1.6,0.25) (-1,0)} -- cycle;

\filldraw[red] (1.25,1) circle (2pt);
\filldraw[red] (2.7,1) circle (2pt);

\draw[Green, very thick] (1.25,1) circle (0.25);
\draw[Green, very thick] (2.7,1) circle (0.25);

\draw [Green, very thick, decorate, decoration={snake, amplitude=.4mm, segment length=2mm, post length=0}] (1.5,1) -- (2.45,1);

\draw[Green] (1.25,0.45) node{\small $\bra{a_2}$};
\draw[Green] (2.7,0.45) node{\small $\ket{a_1}$};

\draw[very thick, CadetBlue] (0,0) arc (-90:90:0.5 and 1);
\end{scope}

\begin{scope}[shift={(9,0)}]

\draw[very thick, gray] plot [smooth, tension=1] coordinates {(2.5,0) (3.1,0.25) (3.5,0.9) };
\draw[very thick, gray] plot [smooth, tension=1] coordinates {(2.5,2) (3.1,1.75) (3.5,1.1) };

\fill[gravity, opacity = 0.5]  plot[domain=180:360, smooth, variable=\x] ({cos(\x) + 1}, {(0.5*sin(\x))*(cos(\x) + 1 )^(2/5) + 2}) -- (2,2) -- (0,2);

\fill[gravity, opacity = 0.5]  plot[domain=180:360, smooth, variable=\x] ({-cos(\x) + 1}, {(0.5*sin(\x))*(cos(\x) + 1 )^(2/5) + 2}) -- (0,2) -- (2,2);

\draw[very thick, gray] (-1,2) -- (0,2);
\draw[very thick, gray] (2,2) -- (2.5,2);
\draw[very thick, CadetBlue] (0,2) -- (2,2);

\draw[very thick, gray] (-1,0) -- (2.5,0);

\draw[very thick, Green, dashed] (2.5,2) arc (90:270:0.5 and 1);
\draw[very thick, Green, dashed] (-1,2) arc (90:270:0.5 and 1);

\draw[very thick, gray] plot [smooth, tension=1] coordinates {(-1,0) (-1.6,0.25) (-2,0.9) };
\draw[very thick, gray] plot [smooth, tension=1] coordinates {(-2,1.1) (-1.6,1.75) (-1,2)};

\draw[very thick, Green] (-1,0) arc (-90:90:0.5 and 1);
\draw[very thick, Green] (2.5,0) arc (-90:90:0.5 and 1);

\filldraw[red] (3.5,1) circle (2pt);
\filldraw[red] (-2,1) circle (2pt);

\draw[Green] (-1,-0.4) node{$\bra{a_2}$};
\draw[Green] (2.5,-0.4) node{$\ket{a_1}$};

\draw[very thick, domain=180:360, smooth, variable=\x, CadetBlue] plot ({cos(\x) + 1}, {(0.5*sin(\x))*(cos(\x) + 1 )^(2/5) + 2});

\draw[very thick, domain=180:360, smooth, variable=\x, CadetBlue, dashed] plot ({- cos(\x) + 1}, {(0.5*sin(\x))*(cos(\x) + 1 )^(2/5) + 2});

\draw[very thick, Green, dashed] (-1,2) arc (90:270:0.5 and 1);

\draw [Green, very thick, decorate, decoration={snake, amplitude=.4mm, segment length=2mm, post length=0}] (-0.5,1) -- (3,1);

\end{scope}

\draw[stealth-stealth] (4.4, 1) -- (6.1, 1) node[midway, below]{\small \eqref{transformation_EFT_EE_app}};

\end{tikzpicture}
\end{center}
\caption{{\it Left}: the replicated path integral to compute the entanglement entropy of the matter theory. The geometry is a cylinder because we consider thermal states. The left side is the curved spacetime, where we have already integrated out the gravitational dynamics. The red dot on the left corresponds to $y = b$, while the one on the right corresponds to the infrared cutoff $y = L$. {\it Right}: The thermal cylinder under the transformation \eqref{transformation_EFT_EE_app}. The two red dots are mapped to infinity. We regulate the path integral cutting two surfaces and imposing the states $\ket{a_1}$ and $\ket{a_2}$. The gravity region on the left drawing is mapped to the blue region on the right drawing.}
\label{fig:TFD_EFT}
\end{figure}
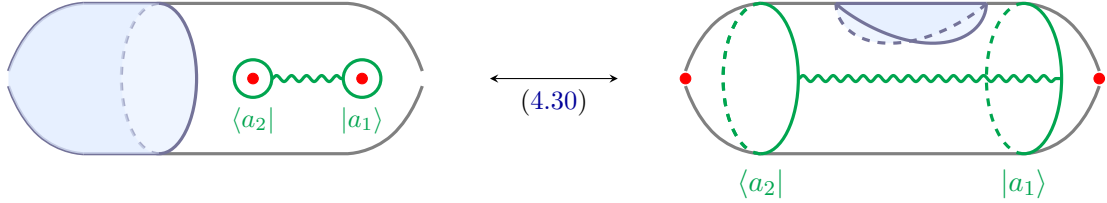

We now turn to the computation of the matter entanglement entropy. We will consider both prescription, the one from Scenario $A$ and the one we are proposing for gravitational theories, Scenario $B$. As anticipated, we want to compute the entanglement entropy of the semi-infinite interval $R = [b, + \infty)$, where $b$ is positive. The replicated partition function for Scenario $A$ is simply
\begin{equation}
    Z_{A, \, n} = \cF^n_{\rm JT} \int [\cD \phi] \, e^{-I_{\rm CFT}(g^{(n)}, \phi)} \ .
    \label{Naive_JT_replicated_Z}
\end{equation}
As expected from the analysis performed in Section \ref{sec:Hawking_rad_EE}, at this level of approximation the Influence Functional does not contribute to the entanglement entropy since it cancels in the combination $\log(Z_n /Z^n)$. We can then consider only the matter partition function. This is both ultraviolet and infrared divergent, which forces us to introduce cutoffs. To cure the IR divergence we regulate the radiation region to be $R = [b, L]$. It is then useful to consider the transformation 
\begin{equation}
    p = \frac{\beta}{2 \pi} \log \left[\frac{e^{\frac{2 \pi  y}{\beta }}-e^{\frac{2 \pi  b}{\beta }}}{e^{\frac{2 \pi  L}{\beta }}-e^{\frac{2 \pi  y}{\beta }}} \right] \ , \qquad \qquad p = \tilde \sigma + i \tilde \tau \ .
    \label{transformation_EFT_EE_app}
\end{equation}
that maps the cylinder into itself, albeit sending the points $y_1 = b$ and $y_2 = L$ to the two opposite ends of the cylinder. On the other hand, to cure the UV divergence of the partition function, we regulate by cutting off the cylinder in the $p$ coordinates at $\tilde \sigma_1$ and $\tilde \sigma_2$ such that $\Delta \tilde \sigma / \beta \gg 1$. Performing these cuts, we have to specify boundary conditions on them, which we can call generically $\ket{a_1}$ and $\ket{a_2}$. 

\begin{figure}[t]
\begin{center}
\begin{tikzpicture}

\begin{scope}[shift={(0,0)}]

\draw[very thick, gray] plot [smooth, tension=1] coordinates {(2.5,0) (3.1,0.25) (3.5,0.9) };
\draw[very thick, gray] plot [smooth, tension=1] coordinates {(2.5,2) (3.1,1.75) (3.5,1.1) };

\draw[very thick, gray] (0,0) -- (2.5,0);
\draw[very thick, gray] (0,2) -- (2.5,2);
\draw[very thick, CadetBlue, name path=D] (0,0) -- (-1,0);
\draw[very thick, CadetBlue, name path=E] (0,2) -- (-1,2);

\draw[very thick, CadetBlue, dashed] (0,2) arc (90:270:0.5 and 1);

\draw[very thick, CadetBlue] plot [smooth, tension=1] coordinates {(-1,0) (-1.6,0.25) (-2,0.9) };
\draw[very thick, CadetBlue] plot [smooth, tension=1] coordinates {(-2,1.1) (-1.6,1.75) (-1,2)};

\fill[gravity, opacity = 0.5]  (0,0) arc (-90:90:0.5 and 1) -- (0,2) -- (-1,2) -- plot [smooth, tension=1] coordinates {(-1,2) (-1.6,1.75) (-2,1.1) } -- plot [smooth, tension=1] coordinates {(-2,0.9) (-1.6,0.25) (-1,0)} -- cycle;
\fill[gravity, opacity = 0.5]  (0,0) arc (270:90:0.5 and 1) -- (0,2) -- (-1,2) -- plot [smooth, tension=1] coordinates {(-1,2) (-1.6,1.75) (-2,1.1) } -- plot [smooth, tension=1] coordinates {(-2,0.9) (-1.6,0.25) (-1,0)} -- cycle;

\filldraw[red] (1.25,1) circle (2pt);

\draw[Green, very thick] (1.25,1) circle (0.25);

\draw [Green, very thick, decorate, decoration={snake, amplitude=.4mm, segment length=2mm, post length=0}] (1.5,1) -- (3.5,1);

\draw[very thick, CadetBlue] (0,0) arc (-90:90:0.5 and 1);
\end{scope}

\begin{scope}[shift={(10,0)}]

\draw[very thick, gray] plot [smooth, tension=1] coordinates {(2.5,0) (3.1,0.25) (3.5,0.9) };
\draw[very thick, gray] plot [smooth, tension=1] coordinates {(2.5,2) (3.1,1.75) (3.5,1.1) };

\draw[very thick, gray] (1,0) -- (2.5,0);
\draw[very thick, gray] (1,2) -- (2.5,2);

\draw[very thick, CadetBlue, name path=D] (1,0) -- (-1,0);
\draw[very thick, CadetBlue, name path=E] (1,2) -- (-1,2);

\draw[very thick, CadetBlue, dashed] (1,2) arc (90:270:0.5 and 1);

\draw[very thick, Green, dashed] (2.5,2) arc (90:270:0.5 and 1);
\draw[very thick, Green, dashed] (-1,2) arc (90:270:0.5 and 1);

\fill[gravity, opacity = 0.5]  (1,0) arc (-90:90:0.5 and 1) -- (1,2) -- (-1,2) -- plot [smooth, tension=1] coordinates {(-1,2) (-1.6,1.75) (-2,1.1) } -- plot [smooth, tension=1] coordinates {(-2,0.9) (-1.6,0.25) (-1,0)} -- cycle;
\fill[gravity, opacity = 0.5]  (1,0) arc (270:90:0.5 and 1) -- (1,2) -- (-1,2) -- plot [smooth, tension=1] coordinates {(-1,2) (-1.6,1.75) (-2,1.1) } -- plot [smooth, tension=1] coordinates {(-2,0.9) (-1.6,0.25) (-1,0)} -- cycle;

\draw[very thick, CadetBlue, name path=D] (1,0) -- (-1,0);
\draw[very thick, CadetBlue, name path=E] (1,2) -- (-1,2);

\draw[very thick, CadetBlue] plot [smooth, tension=1] coordinates {(-1,0) (-1.6,0.25) (-2,0.9) };
\draw[very thick, CadetBlue] plot [smooth, tension=1] coordinates {(-2,1.1) (-1.6,1.75) (-1,2)};

\draw[very thick, CadetBlue] (1,0) arc (-90:90:0.5 and 1);

\draw[very thick, Green] (-1,0) arc (-90:90:0.5 and 1);
\draw[very thick, Green] (2.5,0) arc (-90:90:0.5 and 1);

\filldraw[red] (3.5,1) circle (2pt);

\draw[Green] (-1,-0.4) node{$\bra{a_2}$};
\draw[Green] (2.5,-0.4) node{$\ket{a_1}$};

\draw [Green, very thick, decorate, decoration={snake, amplitude=.4mm, segment length=2mm, post length=0}] (-0.5,1) -- (3,1);

\end{scope}

\draw[stealth-stealth] (4.2, 1) -- (7.3, 1) node[midway, below]{\small \eqref{y_to_pR_transformation} and \eqref{y_to_pM_transformation}};

\end{tikzpicture}
\end{center}
\caption{We consider the transformations \eqref{y_to_pR_transformation} and \eqref{y_to_pM_transformation}, which send the point $b$ to one end of the cylinder, while mapping the boundary of AdS$_2$ into itself. In these coordinates, it is simple to find the replica wormhole solution, since the setup is invariant under shifts in the $\tilde \tau$ coordinates. Therefore, the replica wormhole solution can be found simply rescaling $\beta \to n \beta$. This effectively generates a second twist operator at $y = -b$, but in the evaluation of the partition function the corresponding piece doesn't depend on the position, as we expect for any diffeomorphic invariant theory.}
\label{fig:TFD}
\end{figure}
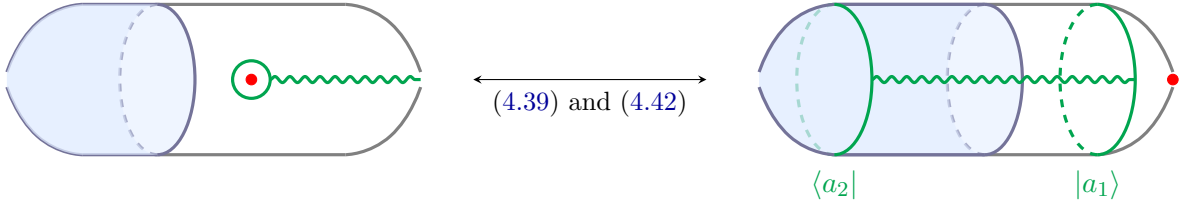

To evaluate the matter partition function, we use the fact that the evolution operator in the limit of $\Delta \tilde \sigma / \beta$ large gets projected to the ground state, 
\begin{equation}
    e^{-\frac{\Delta \tilde \sigma}{n \beta} \, H} = e^{-\frac{\Delta \tilde \sigma}{n \beta} \, E_0}  \ketbra{0} + \dots\ ,
\end{equation}
where corrections are exponentially small in the scaling dimension gap between the vacuum and the first primary operator. Then
\begin{equation}
    \int [\cD \phi] \, e^{-I_{\rm CFT}(g^n, \phi)} = \bra{a_2} e^{-\frac{\Delta \tilde \sigma}{n \beta} \, H} \ket{a_1} = e^{-\frac{\Delta \tilde \sigma}{n \beta} \, E_0}  \braket{a_2}{0} \braket{0}{a_1} \ ,
\end{equation}
where $E_0$ is the ground state for a CFT on the cylinder of unit length, which is
\begin{equation}
    E_0 = -\frac{\pi c}{6} \ .
\end{equation}
Using the inverse transformation $y(p)$ of \eqref{transformation_EFT_EE_app}, one can show that for $\tilde \sigma_1$ large and negative and for $\tilde \sigma_2$ large and positive, one has
\begin{align}
    y(\tilde \sigma_1 + i \tilde \tau) = b + \ve_1 e^{\frac{2 \pi i \tilde \tau}{\beta}} + \dots \quad \qquad  &\text{with}& &\ve_1 = \frac{\beta}{\pi } \, e^{\frac{\pi  (L-b)}{\beta }} \sinh \left(\frac{\pi  (L-b)}{\beta }\right) e^{\frac{2 \pi \tilde \sigma_1}{\beta }} \ ,\\
    y(\tilde \sigma_2 + i \tilde \tau) = L + \ve_2 e^{-\frac{2 \pi i \tilde \tau}{\beta}} + \dots \qquad &\text{with}&   &\ve_2 = \frac{\beta}{\pi } \, e^{\frac{\pi  (b-L)}{\beta }} \sinh \left(\frac{\pi  (L-b)}{\beta }\right) \,  e^{- \frac{2 \pi \tilde \sigma_2}{\beta }}  \ .
\end{align}
This is represented in Figure \ref{fig:TFD}, where two circles at large $|\tilde \sigma|$ are mapped to circles around the twist operators. Setting for simplicity $\ve_1 = \ve_2 \equiv \ve$, we have
\begin{equation}
    \int [\cD \phi] \, e^{-I_{\rm CFT}(g^n, \phi)} = \left[\frac{\beta}{\pi \ve} \sinh(\frac{\pi (L-b)}{\beta}) \right]^{\frac{c}{6n}} \braket{a_2}{0} \braket{0}{a_1} \ ,
\end{equation}
which gives for the matter entanglement entropy 
\begin{equation}
    S_{A} = \frac{c}{3} \log \left[\frac{\beta}{\pi \ve} \sinh(\frac{\pi (L-b)}{\beta}) \right] + \dots \ .
    \label{mat_EE_EFT}
\end{equation}
The dependence on $\ket{a_1}$ and $\ket{a_2}$ is a UV effect that can be absorbed into the definition of $\ve$, see {\it e.g.} \cite{Sully:2020pza}.

The entanglement entropy \eqref{mat_EE_EFT} obtained from the replicated partition function \eqref{Naive_JT_replicated_Z} of Scenario $A$ at first sight may seem a reasonable result to obtain, since it's the universal entanglement entropy for an interval of length $L-b$ in the thermal state. On the other hand, this result becomes suspicious looking at it from a holographic point of view. For instance, let us assume the existence of holographic  Boundary Conformal Field Theories (BCFTs), or possibly Interface CFTs, dual to our JT gravity coupled to a radiation region, in the spirit of \cite{Takayanagi:2011zk, Fujita:2011fp, Almheiri:2019hni, Suzuki:2022xwv, Geng:2022slq, Geng:2022tfc, Afrasiar:2023nir}. The expected thermal entanglement entropy of $[b, +\infty)$ for a BCFT takes the universal form 
\begin{equation}
    S = \frac{c}{6} \log \left[\frac{\beta}{\pi \ve} \sinh(\frac{2\pi b}{\beta}) \right] + \log(g) \ .
    \label{EE_BCFT_thermal}
\end{equation}
where $\log(g)$ is the so-called {\it boundary entropy}. We will now show that the replicated partition function computed following the prescription proposed in Section \ref{sec:Hawking_rad_EE} is able to recover an entanglement entropy consistent with \eqref{EE_BCFT_thermal} and with the island formula.

To follow the proposed gravitational prescription, we consider the replicated geometry of Figure \ref{fig:b_definition}. This has a conical defect at $y = b$. We want to glue to the replicated radiation region a {\it smooth} gravitational region. To perform this gluing, we have to proceed in a similar way as we did before to arrive at the metric \eqref{Original_geometry}. In that case it was possible to chose the saddle $\tau(u) = u$ since it was solving the equations of motion \eqref{non-replicated-EOM}, that also corresponded to having a black hole with constant energy in the gravitational region. We would like to follow the same procedure for the replicated geometry, finding a suitable gluing $\tau(u)$ such that it solves \eqref{non-replicated-EOM} in the replicated geometry. To do it, we have to take into account the fact that the replicated geometry presents a non-trivial stress energy tensor due to the replica boundary conditions. Finding a suitable gluing such that the corresponding EOM \eqref{non-replicated-EOM} are solved is a complicated problem. A solution in the limit $n \to 1$ has been found in \cite{Almheiri:2019qdq}. Here however we want to find a solution for generic $n$. To do it, we take inspiration from Section \ref{sec:Hawking_rad_EE}, and propose a method that in the semiclassical limit is able to perturbatively find the required solution. 

The intuition behind this is that a very simple solution exists for $b \to \infty$. In this case the smooth gravitational geometry would be identical to \eqref{metric_JT_CFT} provided we rescale $\beta \to n \beta$. In that case the gluing would also be straightforward, since our original $\tau(u) = u$ would solve the equations of motion \eqref{non-replicated-EOM}, albeit for the replicated manifold. The key idea is that we can use two conformal transformations similar to \eqref{transformation_EFT_EE_app}, one for the gravitational region and one for the radiation region, to map the point $y = b$ to infinity, while mapping the asymptotic AdS$_2$ boundary into itself. In the new coordinates, finding the replica saddlepoint is much simpler, and going back to the original problem we will find the contribution to the Influence Functional of the smooth gravitational saddle (also called {\it replica wormhole}). 

Let us now find these transformations. Let us start from the flat radiation region, and assume we replicate it $n$ times as in Figure \ref{fig:b_definition}. Then, the transformation 
\begin{equation}
    p_R = \frac{\beta}{2 \pi} \log \left( \frac{\sinh(\frac{\pi}{\beta} (b + y))}{\sinh(\frac{\pi}{\beta} (b - y))} \right)  \ , \qquad \qquad p_R = \tilde \sigma_R + i \tilde \tau_R \ ,
    \label{y_to_pR_transformation}
\end{equation}
maps the point $y = b$ to $p_R \to \infty$. The subscript $R$ indicates that this is the transformation in the radiation region $R$, and thus we should consider only $\tilde \sigma_R >0$. Recall that at $y = b$ (and consequently at $p_R \to \infty$) there is a conical singularity, and that now $\tau$ (and consequently $\tau_R$) is $n \beta$ periodic.

We then consider the following {\it replica wormhole} geometry
\begin{equation}
    \de s^2_{\cM} = \frac{4\pi^2}{n^2 \beta^2} \frac{\de p_{\cM} \, \de \bar p_{\cM}}{\sinh[2](\frac{\pi}{n\beta}(p_{\cM} + \bar p_{\cM}))} \ ,
\end{equation}
where $p_{\cM} = \tilde \sigma_{\cM} + i \tilde \tau_{\cM}$ are coordinates in the gravitational region, and thus we should consider only $\tilde \sigma_{\cM} <0$. Moreover, $\tilde \tau_{\cM}$ is also $n \beta$ periodic. At $\tilde \sigma = 0$, we can then impose the simple gluing
\begin{equation}
    \tilde \tau_{\cM} = \tilde \tau_{R} \ .
\end{equation}
This ensures a smooth replica wormhole solution, and provides a set of global coordinates. In some cases we will then omit the $\cM$ and the $R$ subscripts. We can also find the corresponding transformation between $y$ and $p_{\cM}$ in the gravitational region. The only possibility that ensures a smooth geometry also in the replicated $y$ coordinates is
\begin{equation}
    p_{\cM} = \frac{n \beta}{2 \pi} \log \left( \frac{\sinh(\frac{\pi}{n\beta} (b + y))}{\sinh(\frac{\pi}{n \beta} (b - y))} \right) \ .
    \label{y_to_pM_transformation}
\end{equation}  
Using this transformation we see that in the $y$ coordinates, the replica geometry of the gravitational region remains
\begin{equation}
    \de s^2_{\cM} = \frac{4\pi^2}{n^2 \beta^2} \frac{\de y \, \de \bar y}{\sinh[2](\frac{\pi}{n\beta}(y + \bar y))} \ .
    \label{replica_y_gravitational}
\end{equation}

Finally, the only ingredient missing in order to obtain the (semiclassical) Influence Functional is to understand how the dilaton in the original non-replicated coordinates transforms in the replicated ones. To do it, we find convenient first to look at the non-replicated dilaton solution, and then to obtain the replicated one. First, notice that the transformations \eqref{y_to_pR_transformation} and \eqref{y_to_pM_transformation} coincide in the $n \to 1$ limit. This is the non-replicated geometry. In these coordinates, the original dilaton \eqref{classical_thermal_dilaton} becomes
\begin{equation}
    \Phi = - \frac{2 \pi \Phi_b}{\beta} \left[ \coth(\frac{2 \pi b}{\beta}) \, \coth( \frac{2 \pi \tilde \sigma }{\beta}) - \frac{\cos( \frac{2 \pi \tilde \tau}{\beta}  )}{\sinh(\frac{2 \pi b}{\beta}) \sinh( \frac{2 \pi \tilde \sigma}{\beta} )} \right] \ .
\end{equation}
Close to the boundary, the dilaton does not behave as in \eqref{classical_thermal_dilaton} any longer, but rather takes the form
\begin{equation}
    \Phi \Big|_{\gamma(u)} = \frac{\Phi_b}{\ve} \left[ \coth(\frac{2 \pi b}{\beta})  - \cos( \frac{2 \pi u}{\beta}) \csch(\frac{2 \pi b}{\beta}) \right] \equiv \frac{\Phi_b(u)}{\ve} \ .
    \label{dilaton_bc_p_coordinates}
\end{equation}
These are the boundary conditions for the dilaton in the $p$-coordinates. It is then simple to find the corresponding dilaton solution in the replicated geometry. The choice that maintains the same boundary conditions \eqref{dilaton_bc_p_coordinates} but is also $n \beta$ periodic in $\tau$ is
\begin{equation}
    \Phi_n = - \frac{2 \pi \Phi_b}{n \beta} \left[ \coth(\frac{2 \pi b}{\beta}) \, \coth( \frac{2 \pi \tilde \sigma }{n \beta}) - \frac{\cos( \frac{2 \pi \tilde \tau}{n \beta}  )}{\sinh(\frac{2 \pi b}{\beta}) \sinh( \frac{2 \pi \tilde \sigma}{n \beta} )} \right] \ .
\end{equation}

Now, the gravitational action is simplest evaluated in the $p$ coordinate, with the saddle 
\begin{equation}
    \tilde \sigma = - \ve \ , \qquad \tilde \tau(u) = u \qquad \text{with} \qquad u \in [0, n \beta) \ ,
\end{equation}
giving 
\begin{equation}
    \frac{4}{\kappa^2} \int_0^{n \beta} \de u \, \Phi_n(u) \, \left\{ \tan(\frac{\pi \tilde \tau(u)}{n \beta}) , u \right\}  = \frac{4 \Phi_b }{\kappa^2} \frac{2 \pi^2}{n \beta} \, \coth(\frac{2 \pi b}{\beta}) \ .
\end{equation}
This means that the replicated Influence Functional is\footnote{Notice that for the replica wormhole we still have only $S_0$ since the geometry is still that one of the disk.}, at leading order,
\begin{equation}
    \cF_{n}^B = \exp[S_0 + \frac{8 \pi^2 \Phi_b }{\kappa^2 n \beta} \, \coth(\frac{2 \pi b}{\beta}) ]\ ,
\end{equation}
which crucially depends non-trivially on $n$. Likewise, the open effective theory is
\begin{equation}
    Z_{g, n} = \cF_{n}^B \int [\cD \phi] \, e^{-I_{\rm CFT}( g_n, \phi)} \ .
\end{equation}
Let us then compute the matter entanglement entropy and show that it is now consistent with the island formula. Proceeding as before, we have
\begin{equation}
    \int [\cD \phi] \, e^{-I_{\rm CFT}( g_n, \phi)} = e^{-\frac{\Delta \tilde \sigma}{n \beta} \, E_0}  \braket{a_2}{0} \braket{0}{a_1} + \dots \ .
\end{equation}
Through the inverse transformation \eqref{y_to_pR_transformation}, we have in the $R$ region
\begin{equation}
    y(\tilde \sigma_2 + i \tilde \tau) = b - \ve_2 e^{-\frac{2 \pi i \tilde \tau}{\beta}} + \dots \qquad \text{with} \qquad \ve_2 = \frac{\beta}{\pi} \, \sinh(\frac{2 \pi b}{\beta}) e^{-\frac{2 \pi \tilde \sigma_2}{\beta}}  \ ,
\end{equation}
while through the inverse transformation \eqref{y_to_pM_transformation}, we have in the $\cM$ region
\begin{equation}
    y(\tilde \sigma_1 + i \tilde \tau) = - b + \ve_1 e^{\frac{2 \pi i \tilde \tau}{n \beta}} + \dots \qquad \text{with} \qquad \ve_1 = \frac{n \beta}{\pi} \, \sinh(\frac{2 \pi b}{n \beta}) e^{\frac{2 \pi \tilde \sigma_1}{n \beta}}  \ .
\end{equation}
To compare the cutoffs between the gravity region $\cM$ and the flat region $R$, we impose that the proper lengths are the same. Thus we have
\begin{equation}
    \frac{4\pi^2}{n^2 \beta^2} \frac{\ve_1^2}{\sinh[2](\frac{2\pi b}{n \beta})} = 4 e^{\frac{2 \pi \tilde \sigma_1}{n \beta}} = \ve_2^2 \equiv \ve^2  \ .
    \label{relations_cutoffs}
\end{equation}
Interestingly, this cancels the dependence on the partition function from the contribution coming from the point $y = - b$ in the gravity region, consistent with the fact that the result has to be diffeomorphism invariant on $\cM$. On the other hand, we should be careful in treating the various cutoffs. Let us consider \eqref{relations_cutoffs}, which relates them. In the flat region $R$ we choose a cutoff $\ve_2$, from here on called $\ve$ for simplicity, which is fixed for any number of replicas. On the other hand, in the gravity region $\cM_n$, we notice that if we fix $\tilde \sigma_1$, then the cutoff on $\cM_n$ explicitly depends on the number of replicas $n$. The correct prescription should be that we fix $\ve$ once for all the number of replicas, such that it coincides with the one in the flat region $R$. In turns out that this means that we have to rescale $\tilde \sigma_1 \to n \tilde \sigma_1$, so that all in all we have 
\begin{equation}
    \int [\cD \phi] \, e^{-I_{\rm CFT}( g_n, \phi)} = \left( \frac{2}{\ve} \right)^{\frac{c}{12 n}} \left( \frac{\beta}{\pi \ve} \sinh(\frac{2 \pi b}{\beta}) \right)^{\frac{c}{12 n}} \ .
\end{equation}
The total matter entanglement entropy is then 
\begin{equation}
    S_B = S_0 + \frac{16 \pi^2 \Phi_b }{\kappa^2 \beta} \, \coth(\frac{2 \pi b}{\beta}) + \frac{c}{6} \log(\frac{2}{\ve}) + \frac{c}{6} \log \left[ \frac{\beta}{\pi \ve} \sinh(\frac{2 \pi b}{\beta}) \right] + \dots \ .
\end{equation}
The cutoff in the third term of the RHS can either be absorbed in the definition of $S_0$, or it can be removed rescaling the metric from \eqref{metric_JT_CFT} to the more physical choice
\begin{equation}
    \de s^2_{\cM} = \frac{4\pi^2 \ve^2 }{\beta^2} \frac{\de y \, \de \bar y}{\sinh[2](\frac{\pi}{\beta}(y + \bar y))} \ , \qquad \qquad \de s^2_{R} = \de y \, \de \bar y \ .
    \label{convenient_background}
\end{equation}
Both ways, the final expression for the matter entanglement entropy is 
\begin{equation}
    S_B = S_0 + \frac{16 \pi^2 \Phi_b }{\kappa^2 n \beta} \, \coth(\frac{2 \pi b}{\beta}) + \frac{c}{6} \log(2) + \frac{c}{6} \log \left[ \frac{\beta}{\pi \ve} \sinh(\frac{2 \pi b}{\beta}) \right] + \dots \ ,
    \label{total_mat_EE}
\end{equation}
and in this fashion it is clear that effectively we can say that the the matter fields on the gravitational region contribute to the entanglement entropy with 
\begin{equation}
    S_{g, {\rm mat}} = \frac{c}{6} \log(2) \ ,
\end{equation}
thus of one Hawking pair per mode.

\subsection*{Consistency with the island formula and comments on the information paradox}

We now want to show that \eqref{total_mat_EE} is consistent with the result obtained from the island formula \eqref{Island_Formula}, which we rewrite here for convenience
\begin{equation*}
    S(R) = \operatorname{min}_{I} \left\{ \frac{8 \pi A(\partial I)}{\kappa^2} + S_{\rm bulk \, \, fields} (I \cup R)\right\} \ .
\end{equation*}
The functional over which we minimize is usually called {\it generalized entropy formula} \cite{Almheiri:2019qdq}. In this case the boundary of the island is a point, we call it $y = -a$, in the gravitational region (thus $a$ is positive). Moreover, for lower dimensional dilaton models, the area of the island is represented by the value of the dilaton at the boundary of the island. In formula
\begin{equation}
    \frac{8 \pi A(a)}{\kappa^2} = S_0 + \frac{16 \pi^2 \Phi_b}{\kappa^2\beta} \coth(\frac{2 \pi a}{\beta}) \ .
\end{equation}
Furthermore, we need the entanglement entropy of the matter fields on the fixed curved background \eqref{convenient_background}. Considering the result \eqref{mat_EE_EFT} and the Weyl factors of \eqref{convenient_background}, we have
\begin{equation}
    S_{\rm bulk \, \, fields} (I \cup R) = \frac{c}{6} \log (\frac{2 \beta \sinh[2](\frac{\pi}{\beta}(a+b))}{\pi \ve \sinh(\frac{2 \pi a}{\beta})}) \ .
\end{equation}
The island is found minimizing over $a$ the overall expression
\begin{equation}
    S_{\rm gen}(a,b) = S_0 + \frac{16 \pi^2 \Phi_b}{\kappa^2\beta} \coth(\frac{2 \pi a}{\beta}) + \frac{c}{6} \log (\frac{2 \beta \sinh[2](\frac{\pi}{\beta}(a+b))}{\pi \ve \sinh(\frac{2 \pi a}{\beta})}) \ .
\end{equation}
This procedure results in the condition
\begin{equation}
    \sinh(\frac{2 \pi a}{\beta}) = \frac{96 \pi^2 \Phi_b}{\kappa^2 \beta c} \frac{\sinh(\frac{\pi}{\beta} (a+b))}{\sinh(\frac{\pi}{\beta} (a-b))} \ ,
\end{equation}
and the solution for $b \gtrsim \beta/2\pi$ and $\Phi_b/\beta c \gtrsim 1$ is
\begin{equation}
    a = b + \frac{\beta}{2 \pi} \log(\frac{192 \pi^2 \Phi_b}{\kappa^2 \beta c}) + \dots \ ,
\end{equation}
which implies 
\begin{equation}
    S_{\rm gen} = S_0 + \frac{16 \pi^2 \Phi_b}{\kappa^2\beta} \coth(\frac{2 \pi b}{\beta}) + \frac{c}{6} \log \left[\frac{2 \beta}{\pi \ve} \sinh(\frac{2\pi b}{\beta}) \right] + \dots \ ,
\end{equation}
matching \eqref{total_mat_EE} exactly, and (up to corrections) the BCFT thermal entropy \eqref{EE_BCFT_thermal} setting the boundary entropy
\begin{equation}
    \log(g) = S_0 + \frac{16 \pi^2 \Phi_b}{\kappa^2\beta} + \frac{c}{6} \log (2)
\end{equation}

\begin{figure}[t]
\begin{center}
\begin{tikzpicture}

\draw[-stealth, thick] (-0.5, 0) -- (7, 0);
\draw[-stealth, thick] (0,-0.2) -- (0, 5);

\draw[very thick, red, opacity = 0.2] plot[domain=0:3, smooth, samples=20, variable=\t] ({\t}, {0.5*ln(cosh(3*\t)) + 0.1});
\draw[very thick, red] plot[domain=0:2, smooth, samples=20, variable=\t] ({\t}, {0.5*ln(cosh(3*\t)) + 0.1});

\draw[very thick, red, opacity = 0.2] (0.5,2.75) -- (6.5,2.75);
\draw[very thick, red] (1.98,2.75) -- (6.5,2.75);

\draw[] (6.7,-0.4) node {$t$};
\draw[] (1.98,-0.4) node {$t_P$};
\draw[] (-0.4,4.8) node {$S$};

\draw[] (7.2,2.75) node {$\approx 2 S_0$};

\draw[dashed] (1.99, -0.1) -- (1.99, 4);

\end{tikzpicture}
\end{center}
\caption{The Page curve for the eternal black hole computed using the proposed prescription for the Influence Functional.}
\label{Page_curve}
\end{figure}
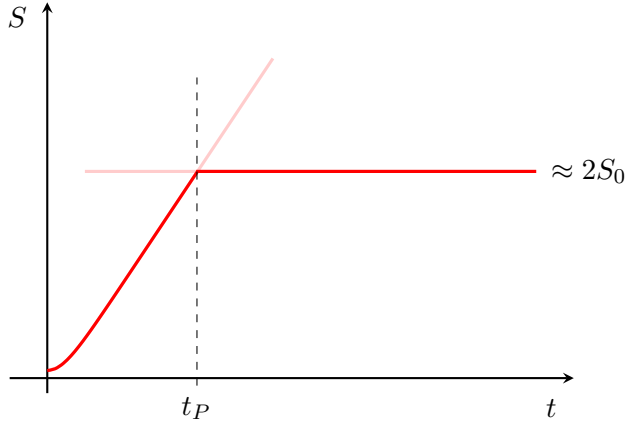

The discussion just presented can also be used to comment on the black hole information paradox. In particular, we can consider the situation in which we replicate the flat region around two points, namely
\begin{align}
    y_1 \, = & \; b + i \tau \ ,\nonumber \\
    y_2 \, = & \; b + i \beta /2 - i \tau \ .
    \label{Page_curve_replica_points}
\end{align}
This is instrumental for the analytic continuation $\tau = - i t$, which allows to consider the out of equilibrium solution. In principle, our definition for the Influence Functional does not allow us to consider out-of-equilibrium states. However, we will argue that replicating the system around \eqref{Page_curve_replica_points}, we will be able to use the results of the previous section at different time scales so that at each instant of time the Influence Functional is well approximated by an equilibrium saddle.

The physical picture in the Lorentzian signature is the one of a two-sided eternal black hole that exchanges quanta with the radiation region $R$, in such a way that the overall energy of the black hole is invariant but entanglement is created between the black hole and $R$. 

Taking inspiration from \cite{Almheiri:2019qdq}, at the beginning of the Lorentzian evolution the dominant saddle to the Influence Functional is given by the disconnected geometry. On the other hand, after the Page time, the dominant contribution is a multi-boundary wormhole solution. In this case, it is not possible to find a suitable conformal transformation that allows to find exactly this wormhole geometry. However, we notice that in the Lorentzian evolution the distance between the two points $y_1$ and $y_2$ increases over time. Therefore, their mutual interaction decreases over time, and the dominant saddle is well approximated at late time by (twice) the one just found for the single replica point \cite{Almheiri:2019qdq}. 

In the case of different saddles, we can compute the moments of the density matrix as
\begin{equation}
    \op{Tr} [\rho^n] = \sum_{\cS_i} \frac{Z_n[\cS_i]}{Z^n[\cS_i]} \ ,
\end{equation}
where $\cS_i$ are the different replica saddles and the summation is over all of them. Therefore, at early and late times, the Rényi entropies are well approximated by
\begin{multline}
    S_n = \frac{1}{1-n} \log \left\{ \left[\frac{\beta}{\pi \ve} \cosh(\frac{2\pi t}{\beta}) \right]^{\frac{c}{6} \left( \frac{1}{n} - n \right)} \right. \\
    \left.+ \, e^{2 S_0 (1-n) + \frac{16 \pi^2 \Phi_b }{\kappa^2\beta} \left( \frac{1}{n} - n \right) \, \coth(\frac{2 \pi b}{\beta})} \left( \frac{2\beta}{\pi \ve} \sinh(\frac{2 \pi b}{\beta}) \right)^{\frac{c}{6} \left( \frac{1}{n} - n \right)} \right\} \ ,
\end{multline}
which implies an entanglement entropy of the form
\begin{equation}
    S = \begin{cases}
    \displaystyle \frac{c}{3} \log \left[\frac{\beta}{\pi \ve} \cosh(\frac{2\pi t}{\beta}) \right] + \dots \qquad & \text{early time} \ ,\\
    \\
    \displaystyle 2 S_0 + \frac{32 \pi^2 \Phi_b}{\kappa^2\beta} \coth(\frac{2 \pi b}{\beta}) + \frac{c}{3} \log \left[\frac{2 \beta}{\pi \ve} \sinh(\frac{2\pi b}{\beta}) \right] + \dots \qquad & \text{late time}  \ .
    \end{cases}
\end{equation}
This produces the well-known Page curve of Figure \ref{Page_curve} \cite{Almheiri:2019yqk}, with a crossover of behaviors around the Page time
\begin{equation}
    t_{P} \approx \frac{3 S_0 \beta}{\pi c} + \dots \ .
\end{equation}

\section{Discussion} \label{sec:Discussion}

In this work we have analyzed holographic theories using the language of open quantum systems, focusing in particular on their Open Effective Field Theory, and on the subtleties arising when considering a replicated partition function to compute Rényi entropies. Motivated by the Lagrangian description of the bulk theory, we have used the Feynman-Vernon Influence Functional approach \cite{Feynman:1963fq} to compute the Open Effective Field Theory of matter fields considering the gravitational field to be an environment. This is tantamount to finding a non-local functional, the Influence Functional, that comes out of path-integrating the environment, and includes all of the effects that the environment -- in our main example, the gravitational theory, but not restricted to this case -- has on the system.
 
At leading order in the system-environment coupling, as usual related to the Newton constant, the Influence Functional is a non-local $T^2$ deformation of the matter theory. This Influence Functional can be used to compute correlation functions between matter fields, and the result is guaranteed to match exactly the prediction using the global theory. On the other hand, we have shown how replicating the partition function to compute Rényi entropies leads to non-trivial possibilities. In summary, the idea revolves around the fact that for gravitational theories, integrating out some degrees of freedom and then replicating the theory does not lead to equivalent results as first replicating the theory and then integrating out some degrees of freedom, as sketched in Figure \ref{non_commutative_diagram}. The reason behind this discrepancy comes from the fact that replicating the theory changes the boundary conditions, and thus also the smooth gravitational background. Therefore, in the former procedure the original non-replicated saddle is being used, while on the latter the replicated saddle is the geometric background of choice. 

One of the main points of this work is to point out that, even though from an EFT point of view the former procedure seems to be natural when computing Rényi entanglement entropies of matter fields, only the latter is able to give answers consistent with unitarity. In particular, based on these suggestions, we have proposed a prescription to compute the gravitational replicated Influence Functional $\cF_n$, and we have shown how this prescription leads to the island formula when computing the matter entanglement entropy. We applied the same formalism to the well known case of a black hole in JT gravity coupled to a CFT$_2$, computing its gravitational Influence Functional and showing how it reproduces the results of the island formula.

At this point, we would like to give some possible interpretations of this result. To begin with, we remark that it is perhaps not a surprise that such a discrepancy appears when computing Rényi entropies, which are more sensitive probes than correlation functions. This in turns gives us many future directions which would be interesting to explore, both for formal and phenomenological developments.

On one hand, the prescription to compute the replicated Influence Functional is forcing us to couple the gravitational (= environment, in this context) degrees of freedom among different replicas. From the point of view of EFTs, where the low energy Hilbert space is assumed to factorize into a graviton sector and a matter sector, namely $\cH = \cH_{\rm grav} \otimes \cH_{\rm mat}$, this is not natural. One possibility is that this assumption is wrong, and that in the UV complete theory it doesn't factorize. There are some examples in which this idea seems to be born out, like in models of black hole evaporation in AdS coupled to a reservoir $R$ which are dual to a Boundary/Interface CFT \cite{Almheiri:2019hni, Rozali:2019day, Geng:2020qvw, Chen:2020uac, Anous:2022wqh}\footnote{For a recent review on the relation between these models and BH evaporation in massless gravity see \cite{Geng:2023qwm}.}. In a Boundary/Interface CFT, the boundary excitations, which would be dual to gravity and low-lying matter fields, are already a complete basis of the BCFT. Indeed, through the Boundary OPE, it is possible to rewrite CFT-bulk operators through CFT-boundary operators, and thus we do not expect to have a factorized Hilbert space. This idea of non-factorization also seems to be realized in generic holographic theories. We can take as a prototypical example the AdS$_3$/CFT$_2$ duality. In this case, the semi-classical geometry is expected to arise from the Virasoro identity block \cite{Hartman:2013mia, Anous:2016kss}, which includes the stress energy tensor and all of its descendants. Then, in this case the factorised Hilbert space in the bulk is a large-$N$ effect.

On the other hand, this explanation doesn't seem to be exhaustive for other examples, at least in this naive way. One case could be SYK coupled to bosonic fields, where the Hilbert space is clearly factorized. In this case, we could study the dynamics of the bosonic fields treating SYK as an environment. The theory is clearly unitary, and at low energy it can be described by the same Schwarzian theory of JT gravity coupled to matter fields. As mentioned in Section \ref{sec:warm_up}, a possible explanation in this case is linked with the underlying averaging needed to obtain the Schwarzian theory. In particular, the entanglement entropy is a non-linear probe of the density matrix, and performing an average at different stages of the calculation leads to different observables. More precisely,
\begin{equation}
    \Tr[\overline \rho_s \log(\overline \rho_s)] \, \neq \, \overline{\Tr[ \rho_s \log(\rho_s)] }\ .
    \label{quench_vs_anneal}
\end{equation}
In Appendix \ref{app:Coarse_grain_EE} we show quantitatively that the RHS in \eqref{quench_vs_anneal} is consistent with the computation performed through Scenario $B$, while the LHS with Scenario $A$. This could have been unexpected {\it a priori}. If the Schwarzian theory is the coarse grained effective theory of SYK, one could expect that only quenched averages can be computed through such effective theory, and that to compute annealed quantities one has to consider a possibly different effective theory for the probes they are interested in. On the other hand, what we have found in this work shows that the same effective theory can be used, albeit with a different set of rules. This is quite a remarkable fact, which suggests that some of the knowledge of the UV theory is preserved through a coarse graining, although only statistically. Work in this direction could be instrumental not only to deepen our understanding of the entanglement structure of strongly chaotic theory, but also for possible experimental realizations \cite{Danshita:2016xbo,Garcia-Alvarez:2016wem,Brzezinska:2022mhj,Uhrich:2023ddx}.

On the other hand, other typical effects of open quantum systems could play an interesting role, on top of all {\it quantum decoherence}. Earlier works studied decoherence properties of CFTs \cite{DelCampo:2019afl}, connections to holography \cite{Ho:2013rra} and to wormholes \cite{Anegawa:2020lzw, Verlinde:2021jwu, Verlinde:2021kgt}. Quantum decoherence takes place when a system enters in contact with a quantum environment made of many degrees of freedom, or a classical system with similar properties. As such the Influence Functional approach of this paper would seem to be perfectly adapted to further investigate decoherence in gravity. This would likely necessitate an understanding of the issues uncovered in this paper on the Schwinger-Keldysh  and related contours.

Finally, these ideas could also be tested on tensor networks models of holography \cite{Swingle:2009bg, Pastawski:2015qua, Chandra:2023dgq, Belin:2023efa} . An interesting recent development is presented in \cite{Belin:2023efa}, where the notion of approximate CFTs is used to develop a random tensor network that could be seen as a discretization of pure AdS$_3$ gravity. The ideas presented there could be implemented by replicating approximate CFTs with light matter operators. We leave this issue for future work.

Let us conclude by remarking that the different entropies considered in this work are not observables in the strict sense of the word. Nevertheless, some versions of Rényi entropies do lend themselves to possible measurement protocols, especially in highly controlled systems where state tomography is available (see. e.g. \cite{Renner:2021qbe}, where this point is also made). It would nevertheless be interesting to identify more conventional observables that are sensitive to the different IF prescriptions developed in this work.

\vspace{1cm}

\noindent {\Large \bf Acknowledgments}

\vspace{0.3cm}
\noindent We would like to thank Tarek Anous, Alex Belin, Nicola Dondi, Jackson Fliss, Marco Meineri and Renato Renner for helpful conversations. We would like to thank Tarek Anous and Marco Meineri for detailed comments on a draft version of this paper. This work has been supported in part by the Fonds National Suisse de la Recherche Scientifique (Schweizerischer Nationalfonds zur Förderung der wissenschaftlichen Forschung) throughProjectGrants200020 182513, the NCCR51NF40-141869 The Mathematics of Physics (SwissMAP), and by the DFG Collaborative Research Center (CRC) 183 Project No. 277101999 - project A03.

\appendix

\section{Green's functions for the {\it n}-sheeted geometry} \label{sec:n-sheeted_Greens_Functions}

\subsection{Two point Green's functions for gapped theories}

In this Section we want to compute the Green's function of the the simplest gapped theory, a free massive scalar in two dimensions, in the $n$-branched cover geometry used in the text. We follow the approach of \cite{Calabrese:2004eu}. We then consider a system with action 
\begin{equation}
	I[\phi] = \int \de^2 x \left( \frac{1}{2} \, \partial_{\mu} \phi \, \partial^{\mu} \phi + \frac{1}{2} \, m^2 \,  \phi^2 \right) \ .
\end{equation}
As we saw in Section \ref{sec:QFT_EE}, the two-point Green's function of the $n$-sheeted geometry is linked to the the entanglement entropy of the system since 
\begin{equation}
	\partial_{m^2} \log Z_n = -\frac{1}{2 Z_n} \int [\cD \phi] \, \phi^2(x) e^{- I_n[\phi] } = -\frac{1}{2} \int \de^2 x \, G_n(x,x) \ .
\end{equation}
Moreover, as discussed previously, this computation greatly simplifies if we exploit the symmetries of the system: we consider polar coordinates and we compute the entanglement entropy of a semi-infinite line extending from the origin to infinity. The $n$-sheeted geometry is then an $n$-cover of the two dimensional plane. To simplify further, we fist consider a disk with radius $r \in [0,L]$ and with angular variable $\theta \in [0, 2\pi n)$, and then we will take the limit $L \to \infty$. This also means that the region for which we will compute the entanglement entropy is $[0, L]$, in the limit $L \to \infty$. This is what has been done in \cite{Calabrese:2004eu}, so we will review their computation and we will extend it to different integrals involving more than one Green's function.

Finding $G_n(x_1, x_2)$ entails finding the orthonormal eigenfunctions that quantise the field $\phi$ in the replicated geometry, namely the eigenfunctions of
\begin{equation}
	(- \square + m^2) \phi(x) = 0 \ ,
\end{equation}
which are given, in polar coordinates, by
\begin{equation}
	\phi^c_{k,i}(r, \theta) = \cos( k \theta/n ) J_{k/n}(\lambda_{i, k/n} r) \ , \qquad \qquad \phi^s_{k,i}(r, \theta) = \sin( k \theta/n ) J_{k/n}(\lambda_{i, k/n} r) \ .
	\label{eigenfunctions_n-sheeted_geometry}
\end{equation}
Here $k \in \mathbb N$ so that the trigonometric functions are $2 \pi n$ periodic, and $\lambda_i$ is chosen so that the eigenfunctions vanish at $r = L$, so
\begin{equation}
	J_{k/n}(\lambda_{i, k/n} L) = 0 \ .
\end{equation}	
Written in the form of \eqref{eigenfunctions_n-sheeted_geometry}, these eigenfunctions are not normalised. The normalisation can be found imposing
\begin{equation}
	\cN_{k,i} \int \de^2 x \, \, \phi^a_{k,i}(r, \theta) \phi^b_{l,j}(r, \theta) = \delta_{k, l} \delta_{i,j} \delta^{a,b} \ ,
\end{equation}
which implies
\begin{equation}
	\cN_{k,i} = \frac{d_k}{2 \pi n L^2} \frac{2}{ J^2_{\frac{k}{n} + 1}(\lambda_{i, k/n} L)} \ .
	\label{normalisation}
\end{equation}
The redial integral can be performed exactly using 
\begin{equation}
	\int_0^1 x \, \de x \, J_{\alpha}(\lambda_{i, \alpha} x) J_{\alpha}(\lambda_{j, \alpha} x) = \frac{1}{2} \, J^2_{\alpha +1 }(\lambda_{i, \alpha}) \, \delta_{i,j} \ .
\end{equation}
In equation \eqref{normalisation} we have also used the same convention of \cite{Calabrese:2004eu} that $d_0 = 1$ and $d_{k>0} = 2$. Using \eqref{eigenfunctions_n-sheeted_geometry} and \eqref{normalisation}, the two point correlator is then
\begin{multline}
	G_n(x_1, x_2) = \sum_{i, k} \cN_{k,i} \frac{\phi_{k,i}(x_1) \phi_{k,i}(x_2)}{\lambda_{i, k/n}^2 L^2 + m^2} \\
	= \frac{1}{2 \pi n} \sum_{k =0}^{\infty} d_k \sum_{i = 1}^{\infty} \frac{2}{L^2  J^2_{\frac{k}{n} + 1}(\lambda_{i, k/n} L) } \frac{J_{k/n}(\lambda_{i, k/n} r_1) J_{k/n}(\lambda_{i, k/n} r_2)}{\lambda_{i, k/n}^2 L^2 + m^2} \, \cC(\theta_1, \theta_2) \ ,
\end{multline}
where
\begin{equation}
	\cC_{k/n}(\theta_1, \theta_2) = \cos\left( \frac{k \theta_1}{n} \right) \cos\left( \frac{k \theta_2}{n} \right) + \sin\left( \frac{k \theta_1}{n} \right) \sin\left( \frac{k \theta_2}{n} \right) \ .
\end{equation}
At this point we can send $L \to \infty$. The variable $\lambda_{i,k}$ then becomes continuous, and the two point function becomes \cite{Calabrese:2004eu}
\begin{equation}
	G_n(x_1, x_2) = \frac{1}{2 \pi n} \sum_{k =0}^{\infty} d_k \int_0^{\infty} \lambda \, \de \lambda \, \frac{J_{k/n}(\lambda r_1) J_{k/n}(\lambda r_2)}{\lambda^2 + m^2}\, \cC_{k/n}(\theta_1, \theta_2) \ ,
\end{equation}

\subsection{Regulating a two point function at coincident points}
Integrating in $\theta$ and $\lambda$, we have 
\begin{equation}
	\int_0^{\infty} \lambda \, \de \lambda \, \frac{J_{k/n}(\lambda r_1) J_{k/n}(\lambda r_2)}{\lambda^2 + m^2} = K_{k/n} (m r_1) \, I_{k/n}(m r_2)  \ , \qquad {\rm for} \, \,\, \, r_1 \geq r_2 \ .
\end{equation}
We remind the reader that to obtain this result one has to take $\theta_{1,2}$ to be $2 \pi n$ periodic. This means that
\begin{equation}
	-\frac{1}{2} \int \de^2 x \, G_n(x, x) = -\frac{1}{2} \sum_{k =0}^{\infty} d_k \int_0^{\infty} r \, \de r \, K_{k/n} (m r) \, I_{k/n}(m r) \ .
\end{equation}
Unfortunately, for all $k$'s, each radial integral is divergent. Exchanging the sum in $k$ and the radial integral we get 
\begin{equation}
	-\frac{1}{2} \int \de^2 x \, G_n(x, x) = - \int_0^{\infty} r \, \de r \, \frac{1}{2} \sum_{k =0}^{\infty} d_k  \, K_{k/n} (m r) \, I_{k/n}(m r) \ .
\end{equation}
However, for all $r$, also the $k$ sum is divergent. To regulate it, we first introduce a regulator $F(x)$ considering the sum
\begin{equation*}
	\frac{1}{2} \sum_{k =0}^{\infty} d_k  \, K_{k/n} (m r) \, I_{k/n}(m r) \, F(k / n \Lambda) \ .
\end{equation*}
We choose $F(x)$ such that $F(0) = 1$, $F^{(j)} = 0$ for all $j \geq 1$, and that vanishes at infinity sufficiently fast. In this way the sum above is convergent. To compute it, we can now use the Euler-MacLaurin (EML) sum formula \cite{Abramowitz}, which states that
\begin{equation}
	\frac{1}{2} \sum_{k = 0}^\infty d_k \, f(k) = \int_0^\infty f(k) \, \de k - \frac{1}{12} \, f^{(1)} (0) - \sum_{j = 2}^\infty \frac{B_{2j}}{(2j)!} \, f^{(2j-1)}(0) \ ,
	\label{EML_formula}
\end{equation}
where $B_{2j}$ are Bernoulli numbers. In our case, the last term in the RHS of \eqref{EML_formula} vanishes since every term of the sum is zero. Then, using the fact that
\begin{equation}
	\partial_k K_{k/n}(mr) \Big|_{k=0} = 0 \ , \qquad \qquad \partial_k I_{k/n}(mr) \Big|_{k=0} = - \frac{1}{n} \, K_{0}(mr) \ ,
\end{equation}
the EML sum formula can be written as
\begin{equation}
	\frac{1}{2} \sum_{k = 0}^\infty d_k \, K_{k/n} (m r) \, I_{k/n}(m r) \, F(k / n \Lambda) = \int_0^\infty \, K_{k/n} (m r) \, I_{k/n}(m r) \, F(k / n \Lambda) \, \de k + \frac{1}{12n} \,  K_{0}^2 (mr) \ .
\end{equation}
Now we can send $\Lambda \to \infty$ to remove the regulator, and we can change variables to $\tilde k = k /n$, so that
\begin{equation}
	\frac{1}{2} \sum_{k = 0}^\infty d_k \, K_{k/n} (m r) \, I_{k/n}(m r) = n \int_0^\infty \, K_{\tilde k} (m r) \, I_{\tilde k}(m r) \, \de \tilde k + \frac{1}{12n} \,  K_{0}^2 (mr) \ .
	\label{EML_without_regulator}
\end{equation}
Now we realise that the first term in the RHS of \eqref{EML_without_regulator} is divergent, but it gets canceled in the combination $\log(Z_n) - n \log(Z)$, so we can neglect it. Last, using the fact that
\begin{equation}
	\int_0^\infty r \, \de r \, K_{0}^2 (mr) = \frac{1}{2 m^2} \ ,
\end{equation}
we obtain
\begin{equation}
	\int \de^2 x \, G_n(x, x) = \frac{1}{12 m^2 n} \ .
\end{equation}
In the main text we have then used this result to obtain
\begin{equation*}
	\partial_{m^2} \log \left( \frac{Z_n}{Z^n} \right) = -\frac{1}{24 m^2} \left( \frac{1}{n} - n \right) \ .
\end{equation*}
which gives
\begin{equation*}
	S = \frac{1}{6} \log \left( \frac{1}{m \, \ve}\right) \ .
\end{equation*}

\subsection{Products of two Green's functions}

In this section we would like to compute integrals of the form
\begin{equation*}
	\int \de^2 x_1 \, \de^2 x_2 \, G_n(x_1,x_2) \, G_n(x_1,x_2) \ .
\end{equation*}
In the previous Section we have already massaged the two point function to the form
\begin{multline}
	G_n(x_1, x_2) = \frac{1}{2 \pi n} \sum_{k =0}^{\infty} d_k \Big[ \theta(r_1 - r_2) \,  K_{k/n}(m r_1) \, I_{k/n}(m r_2) \\
	+ \theta(r_2 - r_1) \,  K_{k/n}(m r_2) \, I_{k/n}(m r_1) \Big] \cC_{k/n}(\theta_1, \theta_2)  \ ,
\end{multline}
where $\theta(x)$ is the Heaviside step-function. Then we can compute
\begin{multline}
	\int \de^2 x_1 \, \de^2 x_2 \, G_n(x_1,x_2) \, G_n(x_1,x_2) = \\
	 \frac{1}{4 \pi^2 n^2} \sum_{k_\alpha, k_\beta =0}^{\infty} d_{k_\alpha} d_{k_\beta} \left[ 2 \int_{0}^{\infty} r_1 \, \de r_1  \int_0^{r_1}  r_2  \, \de r_2  \, K_{k_\alpha/n}(r_1) K_{k_\beta/n}(r_1) I_{k_\alpha/n}(r_2) I_{k_\beta/n}(r_2) \right] \\
	  \times \int_0^{2 \pi n} \de \theta_1  \int_0^{2 \pi n} \de \theta_2 \, \cC_{k_\alpha/n}(\theta_1, \theta_2) \, \cC_{k_\beta/n}(\theta_1, \theta_2) \ .
	  \label{GnGn_explicit_calculation}
\end{multline}
The factor of two in the brackets comes from the symmetry of the integral in the exchange $r_1 \leftrightarrow r_2$. The angular integral evaluates to
\begin{equation}
	\int_0^{2 \pi n} \de \theta_1  \int_0^{2 \pi n} \de \theta_2 \, \cC_{k_\alpha/n}(\theta_1, \theta_2) \, \cC_{k_\beta/n}(\theta_1, \theta_2) =   \begin{cases}
    4 \pi^2 n^2 \, \delta_{k_\alpha , k_\beta}  & \text{for } k_{\alpha} = 0 \ ,\\
    2 \pi^2 n^2 \, \delta_{k_\alpha , k_\beta}  & \text{for } k_{\alpha} > 0 \ ,
  \end{cases}
  \label{angular_propagators_two_point}
\end{equation}
which means that the double sum in $k_\a$ and $k_\b$ becomes a single sum over $k \equiv k_\a = k_\b$. Finally renaming $s_{1,2} = m r_{1,2}$, we have 
\begin{equation}
	\frac{1}{2} \int \de^2 x_1 \, \de^2 x_2 \, G_n(x_1,x_2) \, G_n(x_1,x_2) = \frac{2}{m^4} \int_{0}^{\infty} s_1 \, \de s_1  \int_0^{s_1}  s_2  \, \de s_2 \,  \frac{1}{2} \sum_{k =0}^{\infty} d_{k}  \, K^2_{k/n}(s_1) I^2_{k/n}(s_2)  \ .
\end{equation}
Using again the EML sum formula with the regulator in place we have 
\begin{multline}
	\frac{1}{2} \sum_{k =0}^{\infty} d_{k}  \, K^2_{k/n}(s_1) I^2_{k/n}(s_2) F(k/n \Lambda) = \\
	\int_0^\infty \de k  \, K^2_{k/n}(s_1) I^2_{k/n}(s_2) F(k/n \Lambda) + \frac{1}{6 n} \, K^2_{0} (s_1) K_{0} (s_2) I_{0}(s_2) F(k/n \Lambda) 
	\label{product_two_G_EML}
\end{multline}
Notice that we have omitted the last term in the RHS of \eqref{EML_formula}, since it vanishes. Moreover, the first term in the RHS of \eqref{product_two_G_EML} scales like $n$, so we can omit it in the computation of the entanglement entropy. All in all, removing the regulator, we have 
\begin{equation}
	\frac{1}{2} \int \de^2 x_1 \, \de^2 x_2 \, G_n(x_1,x_2) \, G_n(x_1,x_2) = \frac{1}{3 m^4 n} \int_{0}^{\infty} s_1 \, \de s_1  \int_0^{s_1}  s_2  \, \de s_2 \,  K^2_{0} (s_1) K_{0} (s_2) I_{0}(s_2)  \ .
\end{equation}
It's not difficult to show that
\begin{equation}
	\int_{0}^{\infty} s_1 \, \de s_1  \int_0^{s_1}  s_2  \, \de s_2 \,  K^2_{0} (s_1) K_{0} (s_2) I_{0}(s_2) = \frac{1}{8} \ ,
\end{equation}
so that in the end we have 
\begin{equation}
	\int \de^2 x_1 \, \de^2 x_2 \, G_n(x_1,x_2) \, G_n(x_1,x_2) = \frac{1}{12 m^4 n}   \ .
\end{equation}
The same approach can be used to find
\begin{equation}
	\int_{\tilde \cM}  \de^2 x_1 \, \de^2 x_2 \, G(x_1,x_2) \, G_n(x_1,x_2) = \frac{1}{12 m^4 n}   \ .
\end{equation}
In particular, in this case the integration domain is only over one of the $n$ sheets, and thus the angular integrals are only defined in the domain $[0, 2\pi)$.

\section{Coarse graining and entanglement entropy} \label{app:Coarse_grain_EE}

In this Section we want to consider the effect of a coarse graining in the computation of the entanglement entropy, or more simply in the moments of the density matrix. In particular, we couple a system with a chaotic environment, and compute coarse grained moments of the reduced density matrix. We show that, at the level of the coarse grained EFT, prescription $B$ is able to quantitatively capture the details of the microscopic computation, and in particular it is possible to match the ``replica wormhole'' contributions. 

To be quantitative, we focus on computing the purity. Considering a global pure state $\rho = \ketbra{\psi}$ and the associated system's density matrix $\rho_s = \Tr_e[\rho]$, the main conceptual point is that
\begin{equation}
    \Tr[\rho^2] \neq \Tr[(\overline \rho_s)^2] \neq \overline{\Tr[(\rho_s)^2]} \ .
\end{equation}
We claim that proposal $A$ computes the middle one (quenched average), while proposal $B$ computes the RHS (annealed average).

It is also possible to show a precise matching with prescription $B$. Considering a generic state
\begin{equation}
    \ket{\psi} = \sum_{ij} \psi_{ij} \ket{s_i} \ket{e_j} \ ,  \qquad \qquad \sum_{ij} |\psi_{ij}|^2 = 1 \ .
\end{equation}
Then, the reduced density matrix and the coarse grained version are
\begin{equation}
    \rho_s = \sum_{ijk} \psi_{ij} \psi_{kj}^* \ketbra{s_i}{s_k} = \sum_{ik} C_{ik} \ketbra{s_i}{s_k} \qquad \text{and} \qquad \overline \rho_s = \sum_{i} \overline{C_{ii}}  \ketbra{s_i}{s_i} \ ,
\end{equation}
where we have defined $C_{ik}$ to ease the notation. Notice that $\Tr[\rho_s] = 1$ imposes $\overline{C_{ii}} \sim 1/d_{s}$ , where $d_s$ is the dimension of the Hilbert space of the system. Then the purity of the coarse grained system density matrix is
\begin{equation}
    \Tr[(\overline \rho_s)^2] = \sum_{i} \overline{C_{ii}}^2 \sim \frac{1}{d_s}\ .
\end{equation}
Finally, we have 
\begin{equation}
    \Tr[\rho_s^2] = \sum_{ij} |C_{ij}|^2 \quad \longrightarrow \quad \overline{\Tr[\rho_s^2]} =  \sum_{i} \overline{ |C_{ii}|^2} + \sum_{i \neq j} \overline{|C_{ij}|^2}\ ,
    \label{B_micro}
\end{equation}
where we have divided the sum into two pieces, one where $i = j$ and one where $i \neq j$. We notice that, due to the system-environment interaction, we can estimate the off-diagonal elements of the density matrix as $\overline{|C_{ij}|^2} \sim 1/ (d_s^2 \, d_e)$ , where $d_e = e^{S_e}$ is the dimension of the environment Hilbert space. Then the purity and the Rényi-two entropy become
\begin{equation}
    \overline{\Tr[\rho_s^2]} \sim  \frac{1}{d_s} + e^{-S_e}\ , \qquad \qquad S_2 \sim \min \Big\{ \log(d_s), \, S_e \Big\} \ .
\end{equation}
In a gravitational setup we would then interpret the second contribution of the RHS of \eqref{B_micro} as a replica wormhole contribution. This discussion is similar to Section 2.3 of \cite{Penington:2019kki}, and is not restricted to holographic theories, but to any environment which has both a finite-dimensional Hilbert space and is chaotic enough for which a coarse grained description is suitable.

\bibliographystyle{utphys}
\bibliography{extendedrefs}

\providecommand{\href}[2]{#2}\begingroup\raggedright\begin{thebibliography}{10}

\bibitem{Maldacena:1997re}
J.~M. Maldacena, ``{The Large N limit of superconformal field theories and
  supergravity},'' \href{http://dx.doi.org/10.4310/ATMP.1998.v2.n2.a1}{{\em
  Adv. Theor. Math. Phys.} {\bfseries 2} (1998) 231--252},
  \href{http://arxiv.org/abs/hep-th/9711200}{{\ttfamily arXiv:hep-th/9711200}}.

\bibitem{Almheiri:2019psf}
A.~Almheiri, N.~Engelhardt, D.~Marolf, and H.~Maxfield, ``{The entropy of bulk
  quantum fields and the entanglement wedge of an evaporating black hole},''
  \href{http://dx.doi.org/10.1007/JHEP12(2019)063}{{\em JHEP} {\bfseries 12}
  (2019) 063}, \href{http://arxiv.org/abs/1905.08762}{{\ttfamily
  arXiv:1905.08762 [hep-th]}}.

\bibitem{Penington:2019npb}
G.~Penington, ``{Entanglement Wedge Reconstruction and the Information
  Paradox},'' \href{http://dx.doi.org/10.1007/JHEP09(2020)002}{{\em JHEP}
  {\bfseries 09} (2020) 002}, \href{http://arxiv.org/abs/1905.08255}{{\ttfamily
  arXiv:1905.08255 [hep-th]}}.

\bibitem{Danshita:2016xbo}
I.~Danshita, M.~Hanada, and M.~Tezuka, ``{Creating and probing the
  Sachdev-Ye-Kitaev model with ultracold gases: Towards experimental studies of
  quantum gravity},'' \href{http://dx.doi.org/10.1093/ptep/ptx108}{{\em PTEP}
  {\bfseries 2017} no.~8, (2017) 083I01},
  \href{http://arxiv.org/abs/1606.02454}{{\ttfamily arXiv:1606.02454
  [cond-mat.quant-gas]}}.

\bibitem{Garcia-Alvarez:2016wem}
L.~Garc\'\i{}a-\'Alvarez, I.~L. Egusquiza, L.~Lamata, A.~del Campo, J.~Sonner,
  and E.~Solano, ``{Digital Quantum Simulation of Minimal AdS/CFT},''
  \href{http://dx.doi.org/10.1103/PhysRevLett.119.040501}{{\em Phys. Rev.
  Lett.} {\bfseries 119} no.~4, (2017) 040501},
  \href{http://arxiv.org/abs/1607.08560}{{\ttfamily arXiv:1607.08560
  [quant-ph]}}.

\bibitem{Brzezinska:2022mhj}
M.~Brzezinska, Y.~Guan, O.~V. Yazyev, S.~Sachdev, and A.~Kruchkov,
  ``{Engineering SYK Interactions in Disordered Graphene Flakes under Realistic
  Experimental Conditions},''
  \href{http://dx.doi.org/10.1103/PhysRevLett.131.036503}{{\em Phys. Rev.
  Lett.} {\bfseries 131} no.~3, (2023) 036503},
  \href{http://arxiv.org/abs/2208.01032}{{\ttfamily arXiv:2208.01032
  [cond-mat.str-el]}}.

\bibitem{Uhrich:2023ddx}
P.~Uhrich, S.~Bandyopadhyay, N.~Sauerwein, J.~Sonner, J.-P. Brantut, and
  P.~Hauke, ``{A cavity quantum electrodynamics implementation of the
  Sachdev--Ye--Kitaev model},''
  \href{http://arxiv.org/abs/2303.11343}{{\ttfamily arXiv:2303.11343
  [quant-ph]}}.

\bibitem{Feynman:1963fq}
R.~P. Feynman and F.~L. Vernon, Jr., ``{The Theory of a general quantum system
  interacting with a linear dissipative system},''
  \href{http://dx.doi.org/10.1016/0003-4916(63)90068-X}{{\em Annals Phys.}
  {\bfseries 24} (1963) 118--173}.

\bibitem{CALDEIRA1983587}
A.~Caldeira and A.~Leggett, ``Path integral approach to quantum brownian
  motion,''
  \href{http://dx.doi.org/https://doi.org/10.1016/0378-4371(83)90013-4}{{\em
  Physica A: Statistical Mechanics and its Applications} {\bfseries 121} no.~3,
  (1983) 587--616}.

\bibitem{CALDEIRA1983374}
A.~Caldeira and A.~Leggett, ``Quantum tunnelling in a dissipative system,''
  \href{http://dx.doi.org/https://doi.org/10.1016/0003-4916(83)90202-6}{{\em
  Annals of Physics} {\bfseries 149} no.~2, (1983) 374--456}.

\bibitem{Jana:2020vyx}
C.~Jana, R.~Loganayagam, and M.~Rangamani, ``{Open quantum systems and
  Schwinger-Keldysh holograms},''
  \href{http://dx.doi.org/10.1007/JHEP07(2020)242}{{\em JHEP} {\bfseries 07}
  (2020) 242}, \href{http://arxiv.org/abs/2004.02888}{{\ttfamily
  arXiv:2004.02888 [hep-th]}}.

\bibitem{Loganayagam:2022zmq}
R.~Loganayagam, M.~Rangamani, and J.~Virrueta, ``{Holographic open quantum
  systems: toy models and analytic properties of thermal correlators},''
  \href{http://dx.doi.org/10.1007/JHEP03(2023)153}{{\em JHEP} {\bfseries 03}
  (2023) 153}, \href{http://arxiv.org/abs/2211.07683}{{\ttfamily
  arXiv:2211.07683 [hep-th]}}.

\bibitem{Saad:2019lba}
P.~Saad, S.~H. Shenker, and D.~Stanford, ``{JT gravity as a matrix integral},''
  \href{http://arxiv.org/abs/1903.11115}{{\ttfamily arXiv:1903.11115
  [hep-th]}}.

\bibitem{Altland:2022xqx}
A.~Altland, B.~Post, J.~Sonner, J.~van~der Heijden, and E.~P. Verlinde,
  ``{Quantum chaos in 2D gravity},''
  \href{http://dx.doi.org/10.21468/SciPostPhys.15.2.064}{{\em SciPost Phys.}
  {\bfseries 15} no.~2, (2023) 064},
  \href{http://arxiv.org/abs/2204.07583}{{\ttfamily arXiv:2204.07583
  [hep-th]}}.

\bibitem{Belin:2020hea}
A.~Belin and J.~de~Boer, ``{Random statistics of OPE coefficients and Euclidean
  wormholes},'' \href{http://dx.doi.org/10.1088/1361-6382/ac1082}{{\em Class.
  Quant. Grav.} {\bfseries 38} no.~16, (2021) 164001},
  \href{http://arxiv.org/abs/2006.05499}{{\ttfamily arXiv:2006.05499
  [hep-th]}}.

\bibitem{Chandra:2022bqq}
J.~Chandra, S.~Collier, T.~Hartman, and A.~Maloney, ``{Semiclassical 3D gravity
  as an average of large-c CFTs},''
  \href{http://dx.doi.org/10.1007/JHEP12(2022)069}{{\em JHEP} {\bfseries 12}
  (2022) 069}, \href{http://arxiv.org/abs/2203.06511}{{\ttfamily
  arXiv:2203.06511 [hep-th]}}.

\bibitem{Belin:2023efa}
A.~Belin, J.~de~Boer, D.~L. Jafferis, P.~Nayak, and J.~Sonner, ``{Approximate
  CFTs and Random Tensor Models},''
  \href{http://arxiv.org/abs/2308.03829}{{\ttfamily arXiv:2308.03829
  [hep-th]}}.

\bibitem{Zinn-Justin:572813}
J.~Zinn-Justin,
  \href{http://dx.doi.org/10.1093/acprof:oso/9780198509233.001.0001}{{\em
  {Quantum Field Theory and Critical Phenomena; 4th ed.}}}
\newblock International series of monographs on physics. Clarendon Press,
  Oxford, 2002.
\newblock \url{https://cds.cern.ch/record/572813}.

\bibitem{Calabrese:2004eu}
P.~Calabrese and J.~L. Cardy, ``{Entanglement entropy and quantum field
  theory},'' \href{http://dx.doi.org/10.1088/1742-5468/2004/06/P06002}{{\em J.
  Stat. Mech.} {\bfseries 0406} (2004) P06002},
  \href{http://arxiv.org/abs/hep-th/0405152}{{\ttfamily arXiv:hep-th/0405152}}.

\bibitem{Almheiri:2019qdq}
A.~Almheiri, T.~Hartman, J.~Maldacena, E.~Shaghoulian, and A.~Tajdini,
  ``{Replica Wormholes and the Entropy of Hawking Radiation},''
  \href{http://dx.doi.org/10.1007/JHEP05(2020)013}{{\em JHEP} {\bfseries 05}
  (2020) 013}, \href{http://arxiv.org/abs/1911.12333}{{\ttfamily
  arXiv:1911.12333 [hep-th]}}.

\bibitem{Penington:2019kki}
G.~Penington, S.~H. Shenker, D.~Stanford, and Z.~Yang, ``{Replica wormholes and
  the black hole interior},''
  \href{http://dx.doi.org/10.1007/JHEP03(2022)205}{{\em JHEP} {\bfseries 03}
  (2022) 205}, \href{http://arxiv.org/abs/1911.11977}{{\ttfamily
  arXiv:1911.11977 [hep-th]}}.

\bibitem{Ryu:2006ef}
S.~Ryu and T.~Takayanagi, ``{Aspects of Holographic Entanglement Entropy},''
  \href{http://dx.doi.org/10.1088/1126-6708/2006/08/045}{{\em JHEP} {\bfseries
  08} (2006) 045}, \href{http://arxiv.org/abs/hep-th/0605073}{{\ttfamily
  arXiv:hep-th/0605073}}.

\bibitem{Lewkowycz:2013nqa}
A.~Lewkowycz and J.~Maldacena, ``{Generalized gravitational entropy},''
  \href{http://dx.doi.org/10.1007/JHEP08(2013)090}{{\em JHEP} {\bfseries 08}
  (2013) 090}, \href{http://arxiv.org/abs/1304.4926}{{\ttfamily arXiv:1304.4926
  [hep-th]}}.

\bibitem{Faulkner:2013ana}
T.~Faulkner, A.~Lewkowycz, and J.~Maldacena, ``{Quantum corrections to
  holographic entanglement entropy},''
  \href{http://dx.doi.org/10.1007/JHEP11(2013)074}{{\em JHEP} {\bfseries 11}
  (2013) 074}, \href{http://arxiv.org/abs/1307.2892}{{\ttfamily arXiv:1307.2892
  [hep-th]}}.

\bibitem{Dong:2017xht}
X.~Dong and A.~Lewkowycz, ``{Entropy, Extremality, Euclidean Variations, and
  the Equations of Motion},''
  \href{http://dx.doi.org/10.1007/JHEP01(2018)081}{{\em JHEP} {\bfseries 01}
  (2018) 081}, \href{http://arxiv.org/abs/1705.08453}{{\ttfamily
  arXiv:1705.08453 [hep-th]}}.

\bibitem{Dong:2016fnf}
X.~Dong, ``{The Gravity Dual of Renyi Entropy},''
  \href{http://dx.doi.org/10.1038/ncomms12472}{{\em Nature Commun.} {\bfseries
  7} (2016) 12472}, \href{http://arxiv.org/abs/1601.06788}{{\ttfamily
  arXiv:1601.06788 [hep-th]}}.

\bibitem{Engelhardt:2014gca}
N.~Engelhardt and A.~C. Wall, ``{Quantum Extremal Surfaces: Holographic
  Entanglement Entropy beyond the Classical Regime},''
  \href{http://dx.doi.org/10.1007/JHEP01(2015)073}{{\em JHEP} {\bfseries 01}
  (2015) 073}, \href{http://arxiv.org/abs/1408.3203}{{\ttfamily arXiv:1408.3203
  [hep-th]}}.

\bibitem{Mertens:2022irh}
T.~G. Mertens and G.~J. Turiaci, ``{Solvable models of quantum black holes: a
  review on Jackiw\textendash{}Teitelboim gravity},''
  \href{http://dx.doi.org/10.1007/s41114-023-00046-1}{{\em Living Rev. Rel.}
  {\bfseries 26} no.~1, (2023) 4},
  \href{http://arxiv.org/abs/2210.10846}{{\ttfamily arXiv:2210.10846
  [hep-th]}}.

\bibitem{Sully:2020pza}
J.~Sully, M.~Van~Raamsdonk, and D.~Wakeham, ``{BCFT entanglement entropy at
  large central charge and the black hole interior},''
  \href{http://dx.doi.org/10.1007/JHEP03(2021)167}{{\em JHEP} {\bfseries 03}
  (2021) 167}, \href{http://arxiv.org/abs/2004.13088}{{\ttfamily
  arXiv:2004.13088 [hep-th]}}.

\bibitem{Takayanagi:2011zk}
T.~Takayanagi, ``{Holographic Dual of BCFT},''
  \href{http://dx.doi.org/10.1103/PhysRevLett.107.101602}{{\em Phys. Rev.
  Lett.} {\bfseries 107} (2011) 101602},
  \href{http://arxiv.org/abs/1105.5165}{{\ttfamily arXiv:1105.5165 [hep-th]}}.

\bibitem{Fujita:2011fp}
M.~Fujita, T.~Takayanagi, and E.~Tonni, ``{Aspects of AdS/BCFT},''
  \href{http://dx.doi.org/10.1007/JHEP11(2011)043}{{\em JHEP} {\bfseries 11}
  (2011) 043}, \href{http://arxiv.org/abs/1108.5152}{{\ttfamily arXiv:1108.5152
  [hep-th]}}.

\bibitem{Almheiri:2019hni}
A.~Almheiri, R.~Mahajan, J.~Maldacena, and Y.~Zhao, ``{The Page curve of
  Hawking radiation from semiclassical geometry},''
  \href{http://dx.doi.org/10.1007/JHEP03(2020)149}{{\em JHEP} {\bfseries 03}
  (2020) 149}, \href{http://arxiv.org/abs/1908.10996}{{\ttfamily
  arXiv:1908.10996 [hep-th]}}.

\bibitem{Suzuki:2022xwv}
K.~Suzuki and T.~Takayanagi, ``{BCFT and Islands in two dimensions},''
  \href{http://dx.doi.org/10.1007/JHEP06(2022)095}{{\em JHEP} {\bfseries 06}
  (2022) 095}, \href{http://arxiv.org/abs/2202.08462}{{\ttfamily
  arXiv:2202.08462 [hep-th]}}.

\bibitem{Geng:2022slq}
H.~Geng, A.~Karch, C.~Perez-Pardavila, S.~Raju, L.~Randall, M.~Riojas, and
  S.~Shashi, ``{Jackiw-Teitelboim Gravity from the Karch-Randall Braneworld},''
  \href{http://dx.doi.org/10.1103/PhysRevLett.129.231601}{{\em Phys. Rev.
  Lett.} {\bfseries 129} no.~23, (2022) 231601},
  \href{http://arxiv.org/abs/2206.04695}{{\ttfamily arXiv:2206.04695
  [hep-th]}}.

\bibitem{Geng:2022tfc}
H.~Geng, ``{Aspects of AdS$_{2}$ quantum gravity and the Karch-Randall
  braneworld},'' \href{http://dx.doi.org/10.1007/JHEP09(2022)024}{{\em JHEP}
  {\bfseries 09} (2022) 024}, \href{http://arxiv.org/abs/2206.11277}{{\ttfamily
  arXiv:2206.11277 [hep-th]}}.

\bibitem{Afrasiar:2023nir}
M.~Afrasiar, D.~Basu, A.~Chandra, V.~Raj, and G.~Sengupta, ``{Islands and
  dynamics at the interface},''
  \href{http://arxiv.org/abs/2306.12476}{{\ttfamily arXiv:2306.12476
  [hep-th]}}.

\bibitem{Almheiri:2019yqk}
A.~Almheiri, R.~Mahajan, and J.~Maldacena, ``{Islands outside the horizon},''
  \href{http://arxiv.org/abs/1910.11077}{{\ttfamily arXiv:1910.11077
  [hep-th]}}.

\bibitem{Rozali:2019day}
M.~Rozali, J.~Sully, M.~Van~Raamsdonk, C.~Waddell, and D.~Wakeham,
  ``{Information radiation in BCFT models of black holes},''
  \href{http://dx.doi.org/10.1007/JHEP05(2020)004}{{\em JHEP} {\bfseries 05}
  (2020) 004}, \href{http://arxiv.org/abs/1910.12836}{{\ttfamily
  arXiv:1910.12836 [hep-th]}}.

\bibitem{Geng:2020qvw}
H.~Geng and A.~Karch, ``{Massive islands},''
  \href{http://dx.doi.org/10.1007/JHEP09(2020)121}{{\em JHEP} {\bfseries 09}
  (2020) 121}, \href{http://arxiv.org/abs/2006.02438}{{\ttfamily
  arXiv:2006.02438 [hep-th]}}.

\bibitem{Chen:2020uac}
H.~Z. Chen, R.~C. Myers, D.~Neuenfeld, I.~A. Reyes, and J.~Sandor, ``{Quantum
  Extremal Islands Made Easy, Part I: Entanglement on the Brane},''
  \href{http://dx.doi.org/10.1007/JHEP10(2020)166}{{\em JHEP} {\bfseries 10}
  (2020) 166}, \href{http://arxiv.org/abs/2006.04851}{{\ttfamily
  arXiv:2006.04851 [hep-th]}}.

\bibitem{Anous:2022wqh}
T.~Anous, M.~Meineri, P.~Pelliconi, and J.~Sonner, ``{Sailing past the End of
  the World and discovering the Island},''
  \href{http://dx.doi.org/10.21468/SciPostPhys.13.3.075}{{\em SciPost Phys.}
  {\bfseries 13} no.~3, (2022) 075},
  \href{http://arxiv.org/abs/2202.11718}{{\ttfamily arXiv:2202.11718
  [hep-th]}}.

\bibitem{Geng:2023qwm}
H.~Geng, ``{Revisiting Recent Progress in the Karch-Randall Braneworld},''
  \href{http://arxiv.org/abs/2306.15671}{{\ttfamily arXiv:2306.15671
  [hep-th]}}.

\bibitem{Hartman:2013mia}
T.~Hartman, ``{Entanglement Entropy at Large Central Charge},''
  \href{http://arxiv.org/abs/1303.6955}{{\ttfamily arXiv:1303.6955 [hep-th]}}.

\bibitem{Anous:2016kss}
T.~Anous, T.~Hartman, A.~Rovai, and J.~Sonner, ``{Black Hole Collapse in the
  1/c Expansion},'' \href{http://dx.doi.org/10.1007/JHEP07(2016)123}{{\em JHEP}
  {\bfseries 07} (2016) 123}, \href{http://arxiv.org/abs/1603.04856}{{\ttfamily
  arXiv:1603.04856 [hep-th]}}.

\bibitem{DelCampo:2019afl}
A.~Del~Campo and T.~Takayanagi, ``{Decoherence in Conformal Field Theory},''
  \href{http://dx.doi.org/10.1007/JHEP02(2020)170}{{\em JHEP} {\bfseries 02}
  (2020) 170}, \href{http://arxiv.org/abs/1911.07861}{{\ttfamily
  arXiv:1911.07861 [hep-th]}}.

\bibitem{Ho:2013rra}
S.-H. Ho, W.~Li, F.-L. Lin, and B.~Ning, ``{Quantum Decoherence with
  Holography},'' \href{http://dx.doi.org/10.1007/JHEP01(2014)170}{{\em JHEP}
  {\bfseries 01} (2014) 170}, \href{http://arxiv.org/abs/1309.5855}{{\ttfamily
  arXiv:1309.5855 [hep-th]}}.

\bibitem{Anegawa:2020lzw}
T.~Anegawa, N.~Iizuka, K.~Tamaoka, and T.~Ugajin, ``{Wormholes and holographic
  decoherence},'' \href{http://dx.doi.org/10.1007/JHEP03(2021)214}{{\em JHEP}
  {\bfseries 03} (2021) 214}, \href{http://arxiv.org/abs/2012.03514}{{\ttfamily
  arXiv:2012.03514 [hep-th]}}.

\bibitem{Verlinde:2021jwu}
H.~Verlinde, ``{Deconstructing the Wormhole: Factorization, Entanglement and
  Decoherence},'' \href{http://arxiv.org/abs/2105.02142}{{\ttfamily
  arXiv:2105.02142 [hep-th]}}.

\bibitem{Verlinde:2021kgt}
H.~Verlinde, ``{Wormholes in Quantum Mechanics},''
  \href{http://arxiv.org/abs/2105.02129}{{\ttfamily arXiv:2105.02129
  [hep-th]}}.

\bibitem{Swingle:2009bg}
B.~Swingle, ``{Entanglement Renormalization and Holography},''
  \href{http://dx.doi.org/10.1103/PhysRevD.86.065007}{{\em Phys. Rev. D}
  {\bfseries 86} (2012) 065007},
  \href{http://arxiv.org/abs/0905.1317}{{\ttfamily arXiv:0905.1317
  [cond-mat.str-el]}}.

\bibitem{Pastawski:2015qua}
F.~Pastawski, B.~Yoshida, D.~Harlow, and J.~Preskill, ``{Holographic quantum
  error-correcting codes: Toy models for the bulk/boundary correspondence},''
  \href{http://dx.doi.org/10.1007/JHEP06(2015)149}{{\em JHEP} {\bfseries 06}
  (2015) 149}, \href{http://arxiv.org/abs/1503.06237}{{\ttfamily
  arXiv:1503.06237 [hep-th]}}.

\bibitem{Chandra:2023dgq}
J.~Chandra and T.~Hartman, ``{Toward random tensor networks and holographic
  codes in CFT},'' \href{http://dx.doi.org/10.1007/JHEP05(2023)109}{{\em JHEP}
  {\bfseries 05} (2023) 109}, \href{http://arxiv.org/abs/2302.02446}{{\ttfamily
  arXiv:2302.02446 [hep-th]}}.

\bibitem{Renner:2021qbe}
R.~Renner and J.~Wang, ``{The black hole information puzzle and the quantum de
  Finetti theorem},'' \href{http://arxiv.org/abs/2110.14653}{{\ttfamily
  arXiv:2110.14653 [hep-th]}}.

\bibitem{Abramowitz}
M.~Abramowitz and I.~Stegun, {\em {Handbook of Mathematical Functions}}.
\newblock National Bureau of Standards, Washington DC, 1972.

\end{thebibliography}\endgroup

\end{spacing}
\end{document}